\documentclass[12pt]{article}
\usepackage[utf8]{inputenc}
\usepackage{a4,amsmath,amssymb,epsfig,emlines2}

\title{{\Large{\bf Geometrical Models of the Locally Anisotropic Space-Time}}}

\author{V.\,Balan$^{1}$\,, G.Yu.\,Bogoslovsky$^{2}$\,, S.S.\,Kokarev$^{3}$\,, D.G.\,Pavlov$^{4}$\, \\S.V.\,Siparov$^{5}$\,, N.\,Voicu$^{6}$}

\date{
$^{1}${\small {\em University Politehnica of Bucharest, Romania},\, vladimir.balan@upb.ro}
\\[0.5mm]
$^{2}${\small {\em Skobeltsyn Institute of Nuclear Physics, MSU, Moscow\\[-1.5mm]
Russia},\,       bogoslov@theory.sinp.msu.ru}
\\
$^{3}${\small {\em Institute of Hypercomplex Systems in Geometry and Physics, Fryazino,\\[-1.5mm]
Russia},\, logos-center@mail.ru}
\\
$^{4}${\small {\em  Institute of Hypercomplex Systems in Geometry and Physics, Fryazino,\\[-1.5mm]
Russia},\, geom2004@mail.ru}
\\[0.3mm]
$^{5}${\small {\em State University of Civil Aviation, St.Petersburg,\\[-1.5mm] 
Russia},\, sergey@siparov.ru}
\\[0.3mm]
$^{6}${\small {\em ``Transilvania'' University of Bra\c{s}ov, Romania},\, nico.brinzei@unitbv.ro}
}

\begin{document}

\maketitle

{\small {\noindent{\bf Abstract.} Along with the construction of non-Lorentz-invariant effective field theories, recent studies which are based on geometric models of Finsler space-time become more and more popular. In this respect, the Finslerian approach to the problem of Lorentz symmetry violation is characterized by the fact that the violation of Lorentz symmetry is not accompanied by a violation of relativistic symmetry. That means, in particular, that preservation of relativistic symmetry can be considered as a rigorous criterion of the viability for any non-Lorentz-invariant effective field theory. Although this paper has a review character, it contains (with few exceptions) only those results on Finsler extensions of relativity theory, that were obtained by the authors.\\[2mm]
{\bf Key words\,:}\, Lorentz\,-\,,\, Poincare\,-\, and\, gauge\, symmetry\,\,;\,\, spontaneous symmetry breaking\,\,;\\ alternative gravity theories\,\,;\,\, space-time anisotropy\,\,;\,\, Finsler differential geometry.\\}}

\section{Introduction}
\normalsize
Nowadays, the program of geometrization and algebraization of the fundamental laws of nature which was formulated at the early stage of GR development is still not fulfilled. Every step in realization of this program suggests partial or complete reconsideration of the common notions and of the properties of the corresponding to them physical objects. Many basic concepts of the modern physics and mathematics are expressed in terms of the notion of manifold, which allows possibility of universal arythmetization of the events of the physical world and of the relations between them; the notion of manifold is also a symbiosis of  geometric and algebraic ideas.

Despite the abstract character of the manifolds studied in modern physics and mathematics and of a lot of additional structures which geometrically describe the laws of nature, some of these structures still remain rather conservative. First of all, we mention the manifolds endowed with metrics, while the majority of modern geometrical models deal with the metric tensor as a function on the tangent bundle. In every coordinate chart, the metric tensor field depends on the coordinates of the base in an arbitrary smooth way, and it depends bi-linearly on the coordinates of the fiber. Despite the fact that the most natural generalization of this construction is known to the mathematicians for a very long time within Finsler geometry [1-3] which describes the locally anisotropic spaces, the first viable model of Finsler space-time [4] and the based on it special relativistic theory of locally anisotropic space-time [5,\,6] were promoted not long ago. These works were motivated by the suggested at that time and now popular idea [7] of Lorentz symmetry violation, which means that the ``true'' metric of the flat space-time deviates from the Minkowski metric.

Generally speaking, the discussion on space-time anisotropy needs to clarify first two issues\,: i) why this should be done, i.e. what are its physical premises and ii) what does the suggested anisotropy mean. The second question implies that geometry in mathematics corresponds to the theory of measurements in physics, that is, when we speak of, say, space-time curvature, we presume that it will show itself in measurements. If we have in mind the physical applications of the geometrical constructions, the same must be true for anisotropy. Notice that to speak of the curvature or anisotropy of the empty space is possible only when we don’t deal with experimental science at all, and if we do deal with it, the characteristic scale for the possible applications of the theoretical speculations must be provided. The last means that when the necessity to study the space-time anisotropy occurs, one should suggest its local source.

The answer to the first question is less obvious and is rather vast from the point of view of the analysis of the situation in physics (see [8] and additional reasoning below). The result of this analysis has both general and concrete aspects. The general conclusion is that the gravitation theory, i.e.\! GR, was developed for and successfully applied at the scale of planetary systems. When applied to cosmological (galactic) scales in the way in which this is done now, it demands the introduction of corrections that are 25 times larger than the value of mass of the observable Universe and which are related to the existence of the new (still unknown) substances -- dark matter and dark energy, which were not supposed to be present in the initial theory. Obviously, alongside with their tracking, one should make sure that the theoretical models are valid. These models are\,: the so called simplest scalar used in the expression for the Hilbert-Einstein action, i.e. scalar curvature; the geometry used for the space-time description, i.e. Riemann geometry; and the 4D space-time used for the description of the physical reality itself.

Important observations that make simple sense, have sufficient value and statistical validity, but contradict classical GR, are the rotation curves of spiral galaxies. The attempts to modify the theory in order to describe them in an adequate way based on increase of complexity [9,\,10] or change [11] of the simplest scalar, or on the modification of the metric [12], appeared to be either not consistent enough -- f(R)-theories, or imposed as well to introduce a new unknown scalar field or some new unknown interaction. The phenomenological MOND theory [13] required either an arbitrary change of the dynamics law, or an arbitrary change of the expression of the gravitation force, in order to provide an acceptable description of the phenomena observed at galactic scales. Its covariant generalization [14] also leads to the introduction of the new scalar field.

The concrete consequence of the analysis is that there is a necessity to make the next step and to study the possibility to use a new geometry to interpret the observations. The natural generalizations of Riemann geometry are Finsler and Lagrange geometries, both taking into account the dependence of the metric tensor on direction at the given point. This direction can be global -- which corresponds to one of the geometries constructed on the commutative-associative algebra, namely, to Berwald-Moor geometry. If we use the Berwald-Moor metric to interpret the gravitation theory, there appear a fixed number of stationary global sources of gravitation whose nature is unclear. This direction could be local --
and then the interpretation might correspond to the motion of the local sources of curvature. The last one seems well-grounded, since the common features of the gravitation theory and of electrodynamics from the point of view of Lorentz invariance and of the inverse square law were long ago noticed. The corresponding attempts to generalize the theory with the help of the notion of mass currents were undertaken in [15,\,16], and the common geometrical background of both theories was discussed in [17]. Nevertheless, the gravito-electromagnetism [16] doesn’t seem to be self-consistent enough, because one cannot deal with the gravitation charges in the same way as with electric charges\,: the first are sources of curvature, while the second are not. Instead of the introduction of Lorentz force according to a formal analogy, one should require that, in the case of gravitation, the metric becomes anisotropic. This would lead to the gravitational force dependence on the velocity of the test particle and on the vector field corresponding to the motion of the sources of curvature. The literal meaning of the equivalence principle suggests the same\,: the inertial forces might depend on velocities and have large values, while the usual relativistic corrections interpreted as the force dependence on velocities, are small. In this case the application of the Schwarzschild type solutions to the problems stated at galaxy scale is not appropriate, and cannot be used to describe the spiral galaxies dynamics which is revealed by observations.

Turning back to the motivation of the research which deals with Finsler geometric structures of space-time, one should notice that the whole variety of astrophysical data including the anisotropy of the acceleration of the Universe expansion and the anisotropy of relic radiation, points at the anisotropy of space-time only in an indirect way. The same can be said about the baryonic asymmetry problem, a breaking of the discrete space-time symmetries in weak interactions, the problem of anomalous magnetic moment of muon, etc.  This emphasizes the significance of new results obtained in the two independent experiments which show directly the existence of the space-time anisotropy.

In the first of them [18], the precise atomic interferometer was used to measure the phase shifts of the freely falling atoms. The local Lorentz symmetry break larger than 2 standard errors was found, which means that there exists an anisotropic condensate of  unknown nature, and that, this interacts with the gravitation field in such a way, that the central symmetry of the gravitational potential is broken. Consider now the second experiment.

Recently, at Large Hadronic Collider (LHC) there was found a new phenomenon [19] which is now known as Ridge/CMS-effect (CMS stands for Compact Muon Solenoid which is both the detector and the name of the corresponding research collaboration). One of the features of the effect consists in the following. If the proton-proton collisions with the full energy 7 TeV produce more than 100 particles, the planes corresponding to the tracks of every pair of the produced charged particles are oriented in such a way that a significant part of them has a common cross-line coinciding with the initial protons collision axis. This resembles the situation with the elastic scattering of a moving particle on a particle at rest: due to the momentum conservation (the momentum is equal to the flying particle momentum), all the planes to which the tracks of the two particles belong after scattering, have the common cross-line which coincides with the track of the initial flying particle. But contrary to the elastic scattering on the particle at rest, the total momentum of the colliding protons at LHC is equal to zero. This fact and also the fact that the Ridge/CMS-effect is characteristic only to the high multiplicity events are hard to explain by regular considerations\,: the physical origin of the appearance of the preferred direction coinciding with the protons collision axis when a hundred or more particles are emitted, remains unclear.

High multiplicity events take place in case of the central collision of the initial protons. Then the energy density at the moment of the collision is comparable to the energy density shortly after the Big Bang, when instead of hadrons there was quark-gluon plasma. It is clear that dealing with the high multiplicity events in the proton-proton collisions, one should account for the phase transitions corresponding to the high gauge symmetries violations that are accompanied by the vacuum rearrangement. The condensate appearing during such rearrangement is locally isotropic (Higgs type) only in frames of the usual relativistic theory. In the relativistic theory with Lorentz symmetry violation, or, in other words, in the anisotropic theory of relativity, which will be discussed below, the role of the Higgs condensate is played by the axially symmetric anisotropic fermion-antifermion condensate. Besides, when rapid cooling and hadronization of quark-gluon plasma takes place, an entirely anisotropic three-gluon condensate can appear. On the one hand, quantum-field vacuum, that includes the anisotropic condensate, is the physical carrier of the local anisotropy of space-time, and it can be regarded as an anisotropic quintessence, on the other -- it imparts all the particles the properties of quasi-particles in the crystalline environment. In particular, apart from the rest energy, the particles obtain a rest momentum. With regard to the Ridge/CMS-effect, this means that in the reference frame coinciding with the center of masses of the colliding protons, (relative to the laboratory), the total momentum of the appearing primordial plasma differs from zero and lies on the collision axis (this is due to the anisotropy of the condensate, which arises spontaneously along the collision axis). This is why the correlation of paired tracks in the CMS experiment has turned out such that the planes to which the tracks belong cross mostly on the axis of proton collisions.

Thus, the Ridge/CMS-effect directly demonstrates that in the early Universe there spontaneously emerged the axially symmetric local anisotropy of space-time with a group $DISIM_b(2)$ as an inhomogeneous group of local relativistic symmetry and the corresponding Finsler metric. As for the possibility of spontaneous emergence of the complete local anisotropy of space-time with the Abelian homogeneous group of local relativistic symmetry and the corresponding generalized Finslerian Berwald-Moor metric, the answer to this question will depend on the three-particle correlation function, whose measurement is already planned by the CMS collaboration.

In section 2, we consider the relativistic Finslerian $DISIM_b(2)$--invariant model of a flat space-time with partially broken isotropy in the 3D space. It will be shown that in this model the physical carrier of the anisotropy of flat space-time is axially symmetric neutrino-antineutrino condensate, and the model itself underlies the anisotropic special theory of relativity and admits a natural generalization to the case of curved space-time and the Finslerian extension of GR. The mentioned above Finsler extensions of general relativity necessarily leads to the existence of, at least, one gauge vector field and of its interaction with the conserved current of the rest mass. 
A number of astrophysical effects of this interaction were studied in detail in the framework of the approach proposed by S.V.\,Siparov. It is suggested to model the physical real world by the 8-dimensional phase space-time, one of the coordinates of which appears to have a constant value. The discussion of this approach, of its origination and of the corresponding calculated and observed effects is given in section 3 of this review. 
As to section 2, in addition to the flat space-time with partially broken isotropy of the 3D space, it contains a brief review of a three-parameter family of flat Finsler spaces with entirely broken 3D isotropy and with Abelian three-parametric group of relativistic symmetry. The Abelian group structure of the relativistic symmetry was the starting point for a deeper study of Finsler Berwald-Moor space, which for the four-dimensional case belongs to the specified family. 
In section 4 we consider the geometric, algebraic, and physical aspects of
the commutative associative algebras and Berwald-Moor geometries
of various dimensions associated with them. In recent years, studies of this kind were also conducted within the framework of international cooperation between the Romanian Academy
and the Academy of Sciences of the Russian Federation. In particular, thanks to the work of Romanian geometers led by V.\,Balan, the results concerning the algebraic side of the theory of Berwald-Moor metrics for various dimensions were complemented by the specific results originating from the modern differential geometry of Finsler spaces. Their description in a concentrated form can be found in section 5 of this review.

\section{Relativistically invariant Finslerian spaces with \\local Lorentz symmetry violation}

As it is known, space-time is Riemannian within the framework of GR, and the distribution and motion of matter only determines the local curvature of space-time without affecting the geometry of the tangent spaces. In other words, regardless of the properties of the material medium which fills the Riemannian space-time, any flat tangent space-time remains the space of events of SR, i.e. the Minkowski space with its Lorentz symmetry, which is usually identified with the relativistic symmetry.

However, in recent literature there is an increasing interest in the problem of violation of Lorentz symmetry. Particularly, the string-motivated approach to this problem is widely discussed.

The point is that even if the original unified theory of interactions possesses Lorentz
symmetry up to the most fundamental level, this symmetry can be spontaneously broken due to the emergence of the condensate of vector or tensor field. The appearance of such a condensate, or of a constant classical field on the background of Minkowski space, implies that it can affect the dynamics of the fundamental fields and thereby modify the Standard Model of strong, weak and electromagnetic interactions. Since the constant classical field is transformed
by the passive Lorentz transformations as a Lorentz vector or tensor, its influence on the dynamics of fundamental fields of the Standard Model is described by the introduction of the additional terms representing all possible Lorentz-covariant convolutions of the condensate with the Standard fundamental fields into the Standard Lagrangian. The phenomenological
theory, based on such a Lorentz-covariant modification of the Standard model is called the Standard Model Extension (SME) [20-22]. 

By design, the phenomenological SME theory is not Lorentz-invariant, since its Lagrangian is not invariant under active Lorentz transformations of the fundamental fields against the background of fixed condensate. In addition, in the context of SME, a violation of Lorentz symmetry also involves the violation of relativistic symmetry, since the presence of non-invariant condensate breaks the physical equivalence of the different inertial reference systems.

It should be added that in the low-energy limit of gravitation theories with broken Lorentz and relativistic symmetries, there appears an unlimited number of possibilities to build a variety of effective field theories, each of which could potentially explain at least some of the recently discovered astrophysical phenomena (see, e.g., [23]).

The very existence of the Finsler geometric models of space-time within which a violation of Lorentz symmetry occurs without the violation of relativistic symmetry strongly constrains the possible effective field theories with broken Lorentz symmetry: in order to be viable, such theories, in spite of the presence of Lorentz violation, should have the property of relativistic invariance.

Since only two types of Finsler spaces with broken Lorentz symmetry are relativistic invariant [24], we first consider the Finsler spaces of the first type.

\subsection{The relativistically invariant Finslerian spaces with partially broken 3D isotropy}

The metric of such spaces suggested in [4] has the following form
\begin{equation}\label{1}
ds^2=\left[\frac{(dx_0-\boldsymbol\nu d\boldsymbol x)^2}{dx_0^2-d\boldsymbol x^{\,2}}\right]^r
(dx_0^2-d\boldsymbol x^{\,2})\,.
\end{equation} 
This metric depends on two constant parameters $\,r\,$ and $\,\boldsymbol\nu\,,$ and generalizes the Minkowski metric, where $\,r\,$ determines the spatial anisotropy, characterizing, thus, the degree of deviation of (1) from the Minkowski metric. Instead of the 3-parametric group of rotations of Minkowski space, Finsler spaces (1) can have only an 1-parametric group of rotations around the unit vector $\,\boldsymbol\nu\,,$ which presents a physically preferred direction in the 3D space. The translational symmetry
suffers no change\,: space-time translations preserve metric (1) invariant (in this sense, it is natural to consider the family of spaces (1) as a family of flat Finsler spaces). With regard to the transformations connecting different inertial reference frames, the usual Lorentz boosts conformally modify metric (1). Therefore, they do not belong to a group of isometries of this metric. However, by using them, we can construct such transformations [5] which belong to the group of isometries of metric (1). The corresponding generalized Lorentz transformations (\,generalized Lorentz boosts\,) are as follows
\begin{equation}\label{2}
x'\,^i=D(\boldsymbol v,\boldsymbol\nu )\,R^i_j(\boldsymbol v,\boldsymbol\nu )\,L^j_k(\boldsymbol v)\, x^k\,,
\end{equation}
where $\,\boldsymbol v\,$ stands for the velocities of the moving (primed) reference frames, the matrices $\,L^j_k(\boldsymbol v)\,$ are the usual Lorentz boosts, the matrices $\,R^i_j(\boldsymbol v,\boldsymbol\nu )\,$ are the additional rotations of the spatial axes of the moving systems around vectors $\,[\boldsymbol v\,\boldsymbol\nu ]\,$ at angles
\begin{equation}\nonumber
\varphi=\arccos\left\{ 1-\frac{(1-\sqrt{1-\boldsymbol v^{\,2}/c^2})
[\boldsymbol v\boldsymbol\nu ]^2}{
(1-\boldsymbol v\boldsymbol\nu /c)\boldsymbol v^{\,2}}\right\}
\end{equation} 
corresponding to the relativistic aberration of vector $\,\boldsymbol\nu\,,$ and, finally, the diagonal matrices  
\begin{equation}\nonumber
D(\boldsymbol v,\boldsymbol\nu )=\left(\frac{1-\boldsymbol v\boldsymbol\nu /c}
{\sqrt{1-\boldsymbol v^{\,2}/c^2}}
\right)^rI
\end{equation}
present the additional dilatational transformations of the coordinates of events.

In contrast to the usual Lorentz boosts, the generalized boosts (2) determine a 3-parametric non-compact group with generators $\,X_1\,, X_2\,, X_3\,$. Thus, with inclusion of 1-parameter group of rotations around the preferred direction  $\,\boldsymbol\nu\,$   and 4-parameter translation group, the inhomogeneous group of isometries, or in other words, inhomogeneous group of relativistic symmetry of flat Finsler spaces (1) appears to have 8-parameters. To obtain the simplest representation for its generators, it is enough to send the third spatial axis along  $\,\boldsymbol\nu\,$  and rewrite the transformation (2) in the infinitesimal form. As a result, we come to the following eight generators
\begin{equation}\label{3}
\left.
\begin{array}{rcl}
X_1\!\!&\!\!=\!\!&\!\!-(x^1p_0+x^0p_1)-(x^1p_3-x^3p_1)\,,\\
X_2\!\!&\!\!=\!\!&\!\!-(x^2p_0+x^0p_2)+(x^3p_2-x^2p_3)\,,\\
X_3\!\!&\!\!=\!\!&\!\!-rx^ip_i-(x^3p_0+x^0p_3)\,,\\
R_3\!\!&\!\!=\!\!&\!\!x^2p_1-x^1p_2\,; \qquad \qquad \qquad \qquad \qquad p_i=\partial /\partial
x^i\,.\\
\end{array}
\right.
\end{equation}
According to [5], these generators satisfy the commutation relations
\begin{equation}\label{4}
\left.
\begin{array}{llll}
  [X_1X_2]=0\,, & [R_3X_3]=0\,, && \\
  \left[X_3X_1\right]=X_1 \,, & [R_3X_1]=X_2 \,, && \\
  \left[X_3X_2\right]=X_2\,, & [R_3X_2]=-X_1\,; & &   \vspace*{10pt}\\
  \vspace*{10pt}
  \left[p_i p_j\right]=0\,; &&& \\
  \left[X_1p_0\right]=p_1\,,& [X_2p_0]=p_2\,, & [X_3p_0]=rp_0+p_3\,, &
  [R_3p_0]=0\,, \\
  \left[X_1p_1\right]=p_0+p_3\,,& [X_2p_1]=0\,, & [X_3p_1]=rp_1\,, &
  [R_3p_1]=p_2\,, \\
  \left[X_1p_2\right]=0\,, & [X_2p_2]=p_0+p_3\,, & [X_3p_2]=rp_2\,, &
  [R_3p_2]=-p_1\,, \\
  \left[X_1p_3\right]=-p_1\,, & [X_2p_3]=-p_2\,, & [X_3p_3]=rp_3+p_0\,, &
  [R_3p_3]=0\,. \\
\end{array}
\right.
\end{equation}
This shows that the homogeneous isometry group of flat Finsler spaces with 
partially broken 3D isotropy contains four parameters (\,generators $X_1\,,\, X_2\,,\, X_3\,$
and $\,R_3\,$). It is a subgroup of the 11-parametric Weyl group [25], and it is isomorphic to the corresponding 4-parametric subgroup (\,with generators $X_1\,,\, X_2\,,\, X_3\!\!\mid_{r=0}\,$ and $\,R_3\,$) of the homogeneous Lorentz group. Since the 6-parametric homogeneous Lorentz group does not have any 5-parametric subgroup, while its 4-parametric subgroup is unique up to isomorphisms [26], the passage from Minkowski space to Finsler spaces (1) implies a minimum possible violation of Lorentz symmetry. With this, the relativistic symmetry represented now by the generalized Lorentz boosts (2) remains valid [27].

Here it is worth noting the following. Despite the fact that at $\,r=0\,$ the Finsler metric (1) reduces to Minkowski metric $\,ds^2=dx_0^2-d\boldsymbol x^{2}\,,$ the 3-parametric non-compact transformations (2) that serve as the homogeneous relativistic symmetry transformations for Finsler metric (1) don't reduce to the usual Lorentz boost $\,x'^i=L^i_k(\boldsymbol v) x^k\,$  but reduce to the transformations
\begin{equation}\label{5}
x'^i=R^i_j(\boldsymbol
v,\boldsymbol\nu )L^j_k(\boldsymbol v) x^k
\end{equation}
that differ by additional rotations  $\,x'^i=R^i_k(\boldsymbol v,\boldsymbol\nu ) x^k\,$ of the space axes. These rotations are designed so that if a light-beam in one inertial frame has the direction of $\,\boldsymbol\nu\,,$ then it will have
the same direction in all inertial frames.

Thus, at $\,r=0\,,$ i.e. in frames of the usual SR, the transformations (5) are the alternative to the Lorentz boosts, however, in contrast to the Lorentz boosts, for any value of $\,\boldsymbol\nu\,,$ they present a 3-parameter non-compact subgroup of the 6-parametric homogeneous Lorentz group. As it was noted in [27], in order to realize these transformations physically, it is enough to choose $\,\boldsymbol\nu\,$ as a direction at any star and then perform an arbitrary Lorentz boost, supplementing it with such rotations of spatial axes that in the new reference system, the direction at the star does not change. Taken together, these transformations form the specified subgroup (5) of the 6-parametric homogeneous Lorentz group. As a result, we can say that within the framework of SR, $\,\boldsymbol\nu\,$  has no physical meaning and serves to the relativistically invariant calibration of the directions of spatial axes of inertial frames.

In connection with the last statement, it is necessary to make another important remark concerning the 3-parametric non-compact group of homogeneous transformations (5). If one complements this group by the 1-parametric group of rotations around $\,\boldsymbol\nu\,$ and 4-parametric translation group, the result is an 8-parameter subgroup of the Poincare group, whose generators and Lie algebra have in our basis the forms (3) and (4) assuming that $\,r=0\,.$ For such a group the name $ISIM(2)$ is now used, and its homogeneous 4-parametric subgroup  $\,SIM(2)\,,$ which includes (5) and the rotations around $\,\boldsymbol\nu\,,$ is the basis for the so-called Very Special Relativity (VSR) [28]. According to VSR, $\,SIM(2)\,$ symmetry suggests a more fundamental local  space-time symmetry than the local Lorentz symmetry. In particular, the requirement of $\,SIM(2)\,$ symmetry was sufficient to show [29] that neutrinos may have mass along with the lepton number conservation, and it is important that this result can not be obtained within the framework of Lorentz-invariant approach without introducing sterile neutrinos. However, a significant drawback of the VSR is that $\,\boldsymbol\nu\,$ is regarded only as a phenomenological parameter and VSR can not say anything of its physical nature.

Much more meaningful from the physical point of view is the special relativistic theory of the locally anisotropic space-time [5,\,6], based on Finsler metric (1), which describes a family of flat relativistically invariant spaces of events with partially broken 3D isotropy, and hence with broken Lorentz symmetry. Most of the results obtained under such a theory have been reproduced in [30] using alternative methods (see also [31]). In particular, the inhomogeneous 8-parametric group of relativistic symmetry of metric (1) with its Lie algebra (4) were obtained using the method of continuous deformations of algebra $\,ISIM(2)\,$. As a result, the corresponding symmetry is more frequently called $DISIM_b(2)$ symmetry (\,where $\,b\,$ is the new designation of the parameter $\,r\,$), and the theory itself [5,\,6] is more frequently called General Very Special Relativity (\,GVSR\,). 

\subsubsection{The rest momentum in addition to the rest energy}

In order to modify the usual relativistic mechanics in accordance with the requirement of invariance with respect to $DISIM_b(2),$ it is enough to replace the Minkowski line element $\,ds=\sqrt{dx_0^2-d\boldsymbol x^2\,}$ in the integral of action
\begin{equation}\label{6} 
S=-mc\int\limits_a^b\!ds  
\end{equation}
by the Finsler line element (1). As a result, the Lagrange function corresponding to a relativistic particle in a locally anisotropic space (1), is the following
\begin{equation}\label{7}  
L = -mc^2\left (\frac{1-\boldsymbol v\boldsymbol\nu /c}{\sqrt{1-\boldsymbol v^{\,2}/c^2}
}\right )^r
\sqrt{1-\boldsymbol v^{\,2}/c^2}\,.  
\end{equation}
With this, one can get the expression for the energy $\,E\,$ and momentum $\,\boldsymbol p\,$ of the relativistic
particle [5]\,:
\begin{equation}\label{8}  
E = \frac{mc^2}{\sqrt{1-\boldsymbol v^{\,2}/c^2}}\left (\frac{1-\boldsymbol
v\boldsymbol\nu /c}{
\sqrt{1-\boldsymbol v^{\,2}/c^2}}\right )^r\,\left [1-r+r\,\frac{1-\boldsymbol v^{\,2}
/c^2}{
1-\boldsymbol v\boldsymbol\nu /c}\right ]\,,
\end{equation}
\begin{equation}\label{9}
\boldsymbol p = \frac{mc}{\sqrt{1-\boldsymbol v^{\,2}/c^2}}\left (\frac{1-\boldsymbol v
\boldsymbol\nu /c}{
\sqrt{1-\boldsymbol v^{\,2}/c^2}}\right )^r\,\left [(1-r)\boldsymbol v/c+
r\boldsymbol\nu\,\frac{1-\boldsymbol v^{\,2}/c^2}{
1-\boldsymbol v\boldsymbol\nu /c}\right ]\,. 
\end{equation}
According to (8), the particle energy $E$ reaches its absolute minimum $E=m c^2$ at $\boldsymbol v=0$. As for the momentum $\,\boldsymbol p\,,$ then according to (9), at $\,\boldsymbol v=0\,$ it takes the value $\,\boldsymbol p\,{=}\,m c\,r\,\boldsymbol\nu \,.$ Thus, in the anisotropic space with metric (1), in addition to the rest energy $\,E=m c^2\,,$ any massive particle obtains another observable parameter -- the rest  momentum $\,\boldsymbol p = m c\,r\,\boldsymbol\nu \,.$ Note also that as shown in [5], the 4-momentum  $\,p^i=(\,p^0\!\!=\!\!E/c\,, \,\boldsymbol p\,)\,$ satisfies the $DISIM_b(2)$-invariant dispersion relation, which we give here in the form\,: 
\begin{equation}\label{10} 
(p_0^2-{\boldsymbol
 p}^2) = {(mc)}^2(1-r)^{(1-r)} (1+r)^{(1+r)}\left[ {\frac{(p_0-{\boldsymbol {p\nu}})^2}{p_0^2-{\boldsymbol p}^2}}\right]^{r}\,.   
\end{equation}
In the non-relativistic limit, the Lagrange function (7) has the following form
\begin{equation}\nonumber
L=-mc^2+mcr(\boldsymbol v\boldsymbol\nu )+(1-r)\frac{m{\boldsymbol v}^{\,2}}{2}+r(1-r)\frac{m(\boldsymbol v\boldsymbol\nu )^2}{2}\,.
\end{equation}
Since this expression $\,-mc^2+mcr(\boldsymbol v\boldsymbol\nu )\,$  which is present here is the total derivative over
time, it can be omitted. As a result, we see that the kinetic energy and momentum  
\begin{equation}\nonumber 
T=\frac{1}{2}m_{\alpha\beta}\,v^\alpha v^\beta \,,\qquad
p_\alpha =m_{\alpha\beta}\,v^\beta \,,\ \qquad (\alpha , \beta
=1,2,3)  
\end{equation}
of the non-relativistic particle  in the anisotropic space (1) are determined by the tensor of the inertial mass [24]\,:
\begin{equation}\label{11}
m_{\alpha\beta}=m(1-r)({\delta}_{\alpha\beta}+r{\nu}_\alpha\,{\nu}_
\beta )\,.  
\end{equation}

Let us now rewrite the Finsler metric (1) so that it is expressed through the four-dimensional quantities\,:
\begin{equation}\label{12}
ds=\left[\frac{(dx_0-\boldsymbol\nu d\boldsymbol
x)^2}{dx_0^2-d\boldsymbol x^{2}}\right]^{\!r/2}\!\!\sqrt{dx_0^2-d\boldsymbol
x^2}=\left[\frac{({\nu}_idx^i)^2}{{\eta}_{ik}dx^idx^k}\right]^{\!r/2}\!\!\sqrt{{\eta}_{ik}dx^idx^k}\,.  
\end{equation} 
Since $\,{\boldsymbol\nu}^2=1\,,$ it is clear that here we have
\begin{equation}\nonumber
{\nu}_i=\{1,-\boldsymbol\nu\}\,,\quad\!\! {\eta}_{ik}=diag\{1,-1,-1,-1\}\,,\quad\!\! {\nu}^i=\{1,\boldsymbol\nu\}\,,\quad\!\! {\nu}_i{\nu}^i=0.
\end{equation}

Finally, we present the physical carrier of the anisotropy of the flat space of events (12) and outline a plan for the further development of the theory. In order to do this, we first turn our attention to the unique property of Finsler metric (12). On the one hand, for $\,r=0\,,$ this turns into Minkowski metric, on the other -- at $\,r=1\,,$ it transforms into the total differential $\,ds={\nu}_idx^i\,.$ The latter means that in this case the action (6) does not depend on the shape of the world line connecting the points $\,a\,$ and $\,b\,.$ In other words, the space-time loses such a physical characteristics as spatial extension, and only a temporal duration which represents the absolute time interval $\,ds={\nu}_idx^i\,$ is left. Moreover, according to (11), the inertial masses of all particles also vanish, and $\,{\nu}_i\,$ becomes not the spurionic 
vector field but is transformed into a covariant constant vector field defined on this degenerate (from the metric point of view) space-time manifold. Incidentally, we note that it is on the space-time manifold, and not on the Minkowski space-time, that the massless fundamental fields (for example, those of the Standard Model) are introduced before spontaneous violation of the initial gauge symmetry and before the appearance of
 masses of the initially massless particles. It is clear due to the fact that in the massless world there are no inertial reference systems, with their mandatory attribute -- the reference stick.  

In accordance with (6) and (12), a constant non-zero field $\,r\,$ defines the specific inseparable interaction of the constant spurionic field $\,{\nu}_i\,$ with massive particles. The effect of this interaction is that the particles obtain -- according to (11), the properties of quasi-particles in an axially symmetric crystalline medium. The complex of constant fields containing the scalar field $\,r\,$ and the spurionic field $\,{\nu}_i\,$ is, thus, the physical carrier of the anisotropy of the flat space of events(12). As it turned out, the null-vector spurionic field $\,{\nu}_i\,$ presents a neutrino-antineutrino
condensate constructed out of constant Weyl spinors. Such spinors are an exact solution of the $DISIM_b(2)$-invariant generalized massive Dirac equation [32], whose Lagrangian has the form
\begin{equation}\label{13} 
{\cal L}=\frac{i}{2}\left(\bar\psi{\gamma}^{\mu}{\partial}_{\mu}\psi
-{\partial}_{\mu} \bar\psi{\gamma}^{\mu}\psi\right) -
m\left[\left( \frac{\nu_{\mu} \bar{\psi} \gamma^{\mu}
\psi}{\bar{\psi}\psi} \right)^{2} \right]^{r/2}\bar{\psi}\psi  \,.
\end{equation}
If the constant scalar field $\,r\,$ is set to zero, the $DISIM_b(2)$-invariant generalized massive Dirac equation becomes the standard massive Dirac equation, which does not have any solution in the form of constant spinors. However, the Weyl equations, arising from the standard massless Dirac equation have solutions in the form of constant spinors, and they provide the possibility to build a constant  null-vector spurionic field $\,{\nu}_i\,.$ But physically, it would be unobservable, since at $\,r=0\,$ we come back to the framework of SR\,: the Finsler metric (12) becomes the Minkowski metric, the rest momentum $\,\boldsymbol p = m c\,r\,\boldsymbol\nu \,$ disappears, the tensor of inertial mass (11) ceases to be a tensor and becomes a scalar $\,m\,.$ Accordingly, all the other effects of spatial anisotropy discussed in [33] will be lost.

\subsubsection{On the problem of construction of the Finslerian GR based on the group ${\bf DISIM_b(2)}$}

As noted in the Introduction, the Ridge/CMS-effect which was observed at the LHC, directly
suggests that in the early Universe the axially symmetric anisotropy of space-time spontaneously arose, and it had the $DISIM_b(2)$ group as an inhomogeneous group of local relativistic symmetry and the respective Finsler metric (12). This is the first and most important reason to regard the problem of constructing a Finslerian general relativity based on the group $DISIM_b(2)\,.$ Needless to speak about the complexity of such a problem, especially because it involves the answers to the questions concerning the nature of the dark matter and dark energy. Despite some advances in this direction, this problem is still not completely solved. So, in the end of section 2.1\,, we suggest a possible way the progress on which is likely to lead to the planned purpose.
    
The key point in the generalization of the flat $DISIM_b(2)$-invariant Finsler metric (12) to a Finsler metric, which describes the corresponding curved locally anisotropic space-time is the following. If the constant values on which the metric (12) depends, namely a scalar $r\,,$ the spurion null-vector ${\nu}_i$ and the spurion tensor $\,{\eta}_{ik}=diag\{1,-1,-1,-1\}\,,$ are replaced by the corresponding conventional fields defined on the space-time manifold, i.e. in the metric (12) the substitutions $r\to r(x)\,,\ {\nu}_i\to {\nu}_i(x)\,,\ {\eta}_{ik}\to g_{ik}(x)\,$ are performed, then the result will be the curved Finsler metric of the following form (see [34]) 
\begin{equation}\label{14}
ds=\left[\frac{({\nu}_idx^i)^2}{g_{ik}dx^idx^k}\right]^{\!r/2}\!\!\sqrt{g_{ik}dx^idx^k}\,,
\end{equation}
where\,: $g_{ik}\!=\!g_{ik}(x)$ is the Riemannian metric tensor associated with the gravitational field, $r\!=\!r(x)$ is a scalar field, which characterizes the magnitude of the local space-time
anisotropy and ${\nu}_i\!=\!{\nu}_i(x)$ is a null-vector field that indicates the locally preferred directions in the space-time.

At any point of the curved Finsler space (14), the corresponding flat tangent Finsler space (12) has its own values of the parameters $r$ and $\boldsymbol\nu\,.$ These values are nothing but the values of the fields $r(x)$ and $\boldsymbol\nu (x)$ at the point of tangency.

Obviously, the dynamics of a Finsler space (14) is completely determined by the
dynamics of the interacting fields $\,g_{ik}(x), r(x), {\nu}_i(x),$ and these fields together with
fields of matter form a unified dynamic system. Therefore, in contrast to the existing purely geometric approaches to the Finsler generalization of Einstein's equations, our approach [34] to this problem is based on the use of methods of the conventional theory of interacting fields.

The fact that during the transition from a flat $DISIM_b(2)$-invariant Finsler metric (12) to a curved Finsler metric (14), we replaced the spurion tensor $\,{\eta}_{ik}=diag\{1,-1,-1,-1\}\,$ and the spurion null-vector ${\nu}_i$ by the conventional fields, became the property of metric (14) invariance with regard to the following local transformations
\begin{equation}\label{15}
g_{ik}\ \to\ e^{2\sigma (\,x\,)}\,g_{ik}\,,\qquad 
\nu _i\ \to\ e^{(\,r-1\,)\sigma (\,x\,)\,/\,r}\,\nu _i\,,\qquad  r\ \to\ r\,,
\end{equation}
where $\sigma (x)$ is an arbitrary function.

In addition to metric (14), the local transformations (15) leave invariant all the observables. Therefore, in the theory of gravitation based on the group $DISIM_b(2),$ the transformations (15) have the meaning of local gauge transformations. For example, the action
\begin{equation}\nonumber
S=-\frac{1}{c}\int\mu{^\ast}
\left(\frac{\nu_i\,v^i}{\sqrt{\,g_{ik}\,v^i\,v^k}}\right )^{4r}
\sqrt{-g}
\,\,d^{\,4}x
\end{equation}
for a compressible fluid in a Finsler space (14) is gauge invariant. In this formula, $\mu ^\ast$ is the invariant energy density of the liquid, $v^i=dx^i/ds,$ and $ds$ is metric (14).

In connection with the above-mentioned local gauge invariance, the dynamical system consisting of the fields $\,g_{ik}\,, r\,, \nu _i\,$ and a compressible fluid must be supplemented by two vector gauge fields $A_i$ and $B_i\,,$ that under local transformations (15) are transformed in the corresponding gradient manner. The $A_i$ field for a certain class of problems is a pure gauge field, and the $B_i$ field, whose gauge transformation has the form
\begin{equation}\nonumber 
B_i\ \to\ B_i+b\,[\,(\,r-1\,)\,\sigma (\,x\,)\,/\,r\,]_{;\,i}\ ,
\end{equation}
where $\,b\,$ is a constant with the dimensionality of length, interacts with the conserved rest mass current $\,j^i\,,$ adding the term proportional to $\,B_ij^i\,$ to the full gauge invariant Lagrangian.

\subsection{The relativistically invariant Finslerian spaces with entirely broken 3D isotropy}

In general case, the metric of relativistically invariant Finslerian spaces with entirely broken 3D isotropy [35] is\,:
\begin{gather}
ds= (dx_0-dx_1-dx_2-dx_3)^{(1+r_1+r_2+r_3)/4}
(dx_0-dx_1+dx_2+dx_3)^{(1+r_1-r_2-r_3)/4}\phantom{ddi}\nonumber\\
\phantom{ds=}{} \times  (dx_0+dx_1-dx_2+dx_3)^{(1-r_1+r_2-r_3)/4}
(dx_0+dx_1+dx_2-dx_3)^{(1-r_1-r_2+r_3)/4}\,.\label{16}
\end{gather}
The three parameters ($r_1\,,$ $r_2$ and $r_3$) characterize the anisotropy of spaces (16) and have the following restrictions
\begin{gather*}
1+r_1+r_2+r_3\ge 0,\qquad 1+r_1-r_2-r_3\ge 0,\\
1-r_1+r_2-r_3\ge 0,\qquad 1-r_1-r_2+r_3\ge 0.
\end{gather*}
It should be noted that if $r_1=r_2=r_3=0,$ then the metric (16) becomes the fourth power root of the product of four 1-forms
\begin{gather*}
ds_{_{B-M}}= [\phantom{d}(dx_0-dx_1-dx_2-dx_3)(dx_0-dx_1+dx_2+dx_3)\phantom{dddi}\\
\phantom{d\,ds_{_{B-M}}=}{} \times\!\!
(dx_0+dx_1-dx_2+dx_3)(dx_0+dx_1+dx_2-dx_3)\phantom{d}]^{1/4}\,.
\end{gather*}
Thus, in this particular case, we obtain the well-known Berwald-Moor metric, but written in the basis, which was introduced in [35].

Now consider the group of isometries of flat Finsler spaces (16). The homogeneous 3-parametric non-compact group of isometries, i.e. the group of the relativistic symmetry of space-time (16) appears to be Abelian, and the transformations belonging to such a group have the same meaning as the ordinary Lorentz boosts. The explicit form of these transformations is
\begin{gather}\label{17}
x'_i=DL_{ik}x_k,
\end{gather}
where\,:
\begin{gather*}
D=e^{-(r_1\alpha _1+r_2\alpha _2+r_3\alpha _3)}\,,
\end{gather*}
$L_{ik}$ are the unimodular matrices that are given by the formulas
\begin{gather}\label{18}
L_{ik}=\left (
\begin{array}{rrrr}
\cal A&-\cal B&-\cal C&-\cal D\\
-\cal B&\cal A&\cal D&\cal C\\
-\cal C&\cal D&\cal A&\cal B\\
-\cal D&\cal C&\cal B&\cal A\\
\end{array}
\right ) ,
\\
{\cal A}=\cosh \alpha _1\cosh \alpha _2\cosh \alpha _3+
\sinh \alpha _1\sinh \alpha _2\sinh \alpha _3,\nonumber\\
{\cal B}=\cosh \alpha _1\sinh \alpha _2\sinh \alpha _3+
\sinh \alpha _1\cosh \alpha _2\cosh \alpha _3,\nonumber\\
{\cal C}=\cosh \alpha _1\sinh \alpha _2\cosh \alpha _3+
\sinh \alpha _1\cosh \alpha _2\sinh \alpha _3,\nonumber\\
{\cal D}=\cosh \alpha _1\cosh \alpha _2\sinh \alpha _3+ \sinh
\alpha _1\sinh \alpha _2\cosh \alpha _3,\nonumber
\end{gather}
$\alpha _1$, $\alpha _2$, $\alpha _3$ are the parameters of the group. Along with the parameters $\alpha _i\,,$ the components $v_i=dx_i/dx_0$ of the coordinate velocity of the primed reference frame can also be used as group
parameters. The parameters $v_i$ and $\alpha _i$ are related by
\begin{gather}\nonumber
v_1=(\tanh\alpha _1-\tanh\alpha_2\tanh\alpha _3)/(
1-\tanh\alpha_1\tanh\alpha_2\tanh\alpha_3),\\\nonumber
v_2=(\tanh\alpha _2-\tanh\alpha_1\tanh\alpha _3)/(
1-\tanh\alpha_1\tanh\alpha_2\tanh\alpha_3),\\
\label{19} v_3=(\tanh\alpha _3-\tanh\alpha_1\tanh\alpha _2)/(
1-\tanh\alpha_1\tanh\alpha_2\tanh\alpha_3).
\end{gather}
The reverse relations have the form
\begin{gather}\nonumber
\alpha _1=\frac{1}{4}\ln \frac{(1+v_1-v_2+v_3)(1+v_1+v_2-v_3)}
{(1-v_1-v_2-v_3)(1-v_1+v_2+v_3)},\\\nonumber \alpha
_2=\frac{1}{4}\ln \frac{(1-v_1+v_2+v_3)(1+v_1+v_2-v_3)}
{(1-v_1-v_2-v_3)(1+v_1-v_2+v_3)},\\\label{20} \alpha
_3=\frac{1}{4}\ln \frac{(1-v_1+v_2+v_3)(1+v_1-v_2+v_3)}
{(1-v_1-v_2-v_3)(1+v_1+v_2-v_3)}.
\end{gather}
As for the generators $X_i$ of the homogeneous 3-parametric group of isometries (17) of the space-time (16), they can be represented as follows
\begin{gather*}\nonumber
 X_1=-r_1x_{\alpha}p_{\alpha}-(x_1p_0+
x_0p_1)+(x_2p_3+ x_3p_2),\\\nonumber
 X_2=-r_2x_{\alpha}p_{\alpha}-(x_2p_0+
x_0p_2)+(x_1p_3+
x_3p_1),\\
X_3=-r_3x_{\alpha}p_{\alpha}-(x_3p_0+
x_0p_3)+(x_1p_2+ x_2p_1),
\end{gather*}
where $p_{\alpha}=\partial /\partial x_{\alpha}$ are the generators of the 4-parametric group of translations. Thus, with
inclusion of the latter, a inhomogeneous group of isometries of the entirely anisotropic Finsler space of events (16) is a 7-parametric group. As to its generators, they satisfy the commutation relations
\begin{alignat*}{4}
 & \left[X_i X_j\right]=0,\qquad && \left[p_{\alpha} p_{\beta}\right]=0, && & \nonumber\\
 & \left[X_1p_0\right]=r_1p_0+p_1,\qquad && [X_2p_0]=r_2p_0+p_2,\qquad  && [X_3p_0]=r_3p_0+p_3, &\nonumber\\
 & \left[X_1p_1\right]=r_1p_1+p_0,\qquad && [X_2p_1]=r_2p_1-p_3,\qquad && [X_3p_1]=r_3p_1-p_2, & \nonumber\\
 & \left[X_1p_2\right]=r_1p_2-p_3,\qquad & & [X_2p_2]=r_2p_2+p_0, \qquad && [X_3p_2]=r_3p_2-p_1, &\nonumber\\
 & \left[X_1p_3\right]=r_1p_3-p_2, \qquad && [X_2p_3]=r_2p_3-p_1, \qquad && [X_3p_3]=r_3p_3+p_0. & 
\end{alignat*}

\section{Modeling real world by the phase space-time and physical results obtained on this way}

The theory and results briefly given below are discussed in detail in the monograph [71].

Let $M = {\mathbb{R}}^4$  be a differentiable 4-dimensional manifold of class $C^{\infty}.$ Let $TM$ be its tangent bundle with coordinates $(x,y) = (x^i, y^i); \ i = 0,1,2,3.$ If $c$ is a parametrizable curve on $M,\ c\!:\![a,b] \rightarrow M,\, t \rightarrow (x^i(t)),$ then its natural extension on $TM$ is $c\!:\![a,b] \rightarrow TM,\, t \rightarrow (x^i(t), y^i(t)),$ where $y^i=dx^i/dt.$ The arc length $s,$ usually chosen as the natural parameter on the curve is thus equal to $s=\int_{0}^{t}\sqrt{g_{ij}y^iy^j}d\tau\,;\ i,\,j=0,\,1,\,2,\,3.$ Suppose that the metric introduced above depends on $y$, i.e. $g_{ij} = g_{ij}(x,y).$ In general, this metric corresponds to the generalized Lagrange geometry, $g_{ij}(x,y)$ is a twice covariant symmetric tensor on $TM$ with the only restrictions: a) $det(g_{ij})\ne 0$ for any $(x,y)$ on $TM,$ and b) when the coordinates on $TM$ change in the way corresponding to the change of coordinates on $M,$ the components of the metric vary in the same way as the components of the (0,2)-tensor on the main manifold $M.$ This means that $TM$ is an 8-dimensional Riemannian manifold, analogous to the 6-dimensional phase space well-known in physics. Its geometry is quite complicated and uses such concepts as nonlinear connection (Ehresmann). But if we limit ourselves to the case of linear coordinate transformations with constant coefficients and of weak gravitational field, i.e. $g_{ij}(x,y)= {\eta}_{ij}+{\varepsilon}_{ij}(x,y)\,;\ {\eta}_{ij}=diag\{1,\,-1,\,-1,\,-1\}\,;\ {\varepsilon}_{ij}(x,y)= \chi {\tilde{\varepsilon}}_{ij}(x,y)\,;\ \chi\ll 1,$ the geometry essentially simplifies, and the definition of $y^i$ makes it possible to use the Sasaki lift for raising and lowering indices on the vertical and horizontal components of the bundle, that is use the same metric tensor. The tensor $g_{ij}$ is a zero-order homogeneous in $y$ tensor, i.e. the metric depends only on the direction of $y,$ but not on its value. This is expressed by the relation $(\partial g_{ij}/\partial y^k)y^k=0.$ If there also holds the condition  $(\partial g_{ij}/\partial y^k)y^j=0,$ then this metric becomes the usual Finsler one [1], but in this approach this is not assumed.

The described formalism means that alongside with the use of a new geometry for the modeling of phenomena in the physical world, instead of the space and time of Newton or of the Minkowski space-time, the 8-dimensional phase space-time is introduced. The character of its extra dimensions is not formal, but they have clear physical meaning, due to the used approach. Clearly, the correspondence between the Lagrangian and Hamiltonian formalism now obtains a new dimension. It should be noted that similarly to the situation when the transition from Newton's time and space to the Minkowski space-time took place and the fundamental constant $c$ with the dimension of speed was demanded, 
the transition from Minkowski space-time to the 8-dimensional phase space-time demands another fundamental constant, $l,$ this time with the dimension of length. One can associate it with the fundamental speed and take $l=c/H,$ then $H$ will be a new constant which has the dimension $s^{-1}.$ This suggests that in the interpretations, the following correspondence $(x^0,x^1,x^2,x^3,y^0,y^1,y^2,y^3)\longleftrightarrow (ct,x,y,z,c/H,v_x/H,v_y/H,v_z/H)$ should be borne in mind. One should also pay attention to the fact that all the events would take place in the 7-dimensional subspace of the 8-dimensional phase space-time, one of the coordinates of which is constant according to construction. The symmetry groups corresponding to this space will be the generalized Lorentz group and de Sitter group. The last can be contracted and be used in the Carroll space and in the Newton-Hooke space that are of interest for the astronomical applications. The possibility of separating the resulting space into such parts as $(x,y,z)$ and $(ct,c/H,v_x/H,v_y/H,v_z/H)$ allows the use of Lobachevsky geometry to describe the space of velocities, this geometry was previously used only in the theory of high-energy particles.

Preserving only linear terms proportional to ${\partial{\varepsilon}_{ij}}/{\partial x^k}\,,\ {\partial{\varepsilon}_{ij}}/{\partial y^k}$ and ${{\partial}^{\,2}{\varepsilon}_{ij}}/{\partial x^k\partial y^l}\,,$ one can obtain the generalized geodesics similarly to [36] in the following form
\begin{equation}\label{21}
\frac{dy^i}{ds}+\left({{\Gamma}^i}_{lk}+\frac{1}{2}\eta ^{it}\frac{{\partial}^2{\varepsilon}_{kl}}{\partial x^j\partial y^t}y^j\right)y^ky^l=0\,,
\end{equation}
where ${{\Gamma}^i}_{jk}=\dfrac{1}{2}\eta ^{ih}({\partial{\varepsilon}_{hj}}/{\partial x^k}+{\partial{\varepsilon}_{hk}}/{\partial x^j}-{\partial{\varepsilon}_{jk}}/{\partial x^h})$ is the Christoffel symbol depending on $y.$ Thus, in order to obtain the equations of motion (dynamic equations) in the weak field limit in the anisotropic space, one should use (21), but not
geodesic equation $dy^i/ds+{{\Gamma}^i}_{lk}y^ly^k=0\,,$ which is appropriate in the same approximation only in a space with Riemann geometry. As a result, after certain simplifications and extraction of the anti-symmetric part of the auxiliary tensor introduced in [8], the equation of motion obtained from the geodesic (21) and applied to the spatial cross-section of the space takes the form
\begin{equation}\label{22}
\frac{d\boldsymbol v}{dt}=\frac{c^2}{2}\left\{-\nabla{\varepsilon}_{00}+\left[\boldsymbol v,rot\frac{\partial{\varepsilon}_{00}}{\partial\boldsymbol v}\right]+\nabla (\boldsymbol v,\frac{\partial{\varepsilon}_{00}}{\partial\boldsymbol v})\right\}\,,
\end{equation}
where ${\varepsilon}_{00}$ is the only (temporal) component of the metric tensor, which remains in
the equation of motion in the approximation of the weak field. Regarding (22) as the equation of dynamics, we obtain the expression for the generalized gravitational force depending on velocities [8]
\begin{equation}\label{23}
{\boldsymbol F}^{(g)}= \frac{mc^2}{2}\left\{-\nabla{\varepsilon}_{00}+\left[\boldsymbol v,rot\frac{\partial{\varepsilon}_{00}}{\partial\boldsymbol v}\right]+\nabla \left(\boldsymbol v,\frac{\partial{\varepsilon}_{00}}{\partial\boldsymbol v}\right)\right\}\,.
\end{equation}
The last two equations are obtained from the geodesics corresponding to the field equations for an anisotropic metric. They do not require a special choice of the energy-momentum tensor, and any additional a priori assumptions. The field equations in the anisotropic space in the linear approximation for weak fields retain their form [37], although their terms may now depend on $y$.

To study the dynamics of spiral galaxies, one could choose
\begin{equation}\nonumber
\boldsymbol u\equiv\frac{c^2}{4}\frac{\partial{\varepsilon}_{00}}{\partial\boldsymbol v}\equiv [\boldsymbol\Omega ,\boldsymbol r]\,,
\end{equation}
where
\begin{equation}\nonumber
\boldsymbol\Omega =\frac{c^2}{4}rot\frac{\partial{\varepsilon}_{00}}{\partial\boldsymbol v}
\end{equation}
and
\begin{equation}\nonumber
\boldsymbol\Omega =rot\left(\int\frac{j^{(m)}(r)}{\mid r-r_0\mid}dV\right)\,,
\end{equation}
where $j^{(m)}(r)$ is the mass current density, and $r_0$ corresponds to the observer. Then the equation for the gravitational force obtains the form
\begin{equation}\label{24}
{\boldsymbol F}^{(g)}= \frac{mc^2}{2}\nabla\left\{-{\varepsilon}_{00}+\frac{2}{c^2}\cdot 4(\boldsymbol u,\boldsymbol v)\right\}\,.
\end{equation}
If we demand the existence of limit transition to the usual GRT, then
\begin{equation}\label{25}
{\boldsymbol F}^{(g)}= \frac{mc^2}{2}\nabla\left\{-\sum_n\frac{r_{n,s}}{r_n}+\frac{2}{c^2}\cdot 4(\boldsymbol u,\boldsymbol v)\right\}\,, 
\end{equation}
and it can be shown that the second term under the gradient has the same order of magnitude as the first one at distances of the order of a galaxy radius. It is this that prevents the vanishing of orbital velocity required by the general relativity. At the same time, the motion in the galactic plane and perpendicular to this plane is now described by the different laws, which removes the well-known paradox [38] in the observations of motion of stellar globular clusters.

In the framework of the suggested approach -- anisotropic geometrodynamics (AGD) -- the notion of a point mass is not sufficient to model the elementary (effective) source of gravitation, and one should use a system of ``center plus current'' which represents a gravitational analogue of the circular coil with current around the central charge. The use of such a system for simulation of a spiral galaxy, leads to the expression $v_{orb}\sim const$ for the orbital velocity corresponding to the observed flat rotation curve, and to the empirical Tully-Fisher law $v_{orb}\sim {L_{lum}}^{1/4},$    which has no explanation in general relativity. The same model can explain the observed substantial excess of deflection in some gravitational lenses over the theoretical calculations, which appears to be due to the internal motions of the masses in the galaxy-lens. It has been also shown that in addition to the known convex gravitational lenses, in the AGD there exist concave gravitational lenses. This can lead to the incorrect determination of distances to the sources used as ``standard candles''. And this can account for another interpretation of the data which led to the idea of the acceleration of the Universe expansion and to the notion of dark energy.
 
Calculation of the explosion of the central body in the ``center plus
 current'' model, resulting in the release of the two equal masses in
opposite directions in the plane of the coil leads to trajectories that resemble the well-known observations obtained by the HUBBLE telescope
(compare Fig.\,1\,-right and Fig.\,1\,-left). 

\begin{figure}[htbp]
\begin{center}
 \vspace*{-4.5mm}
  {
  \epsfig
{file=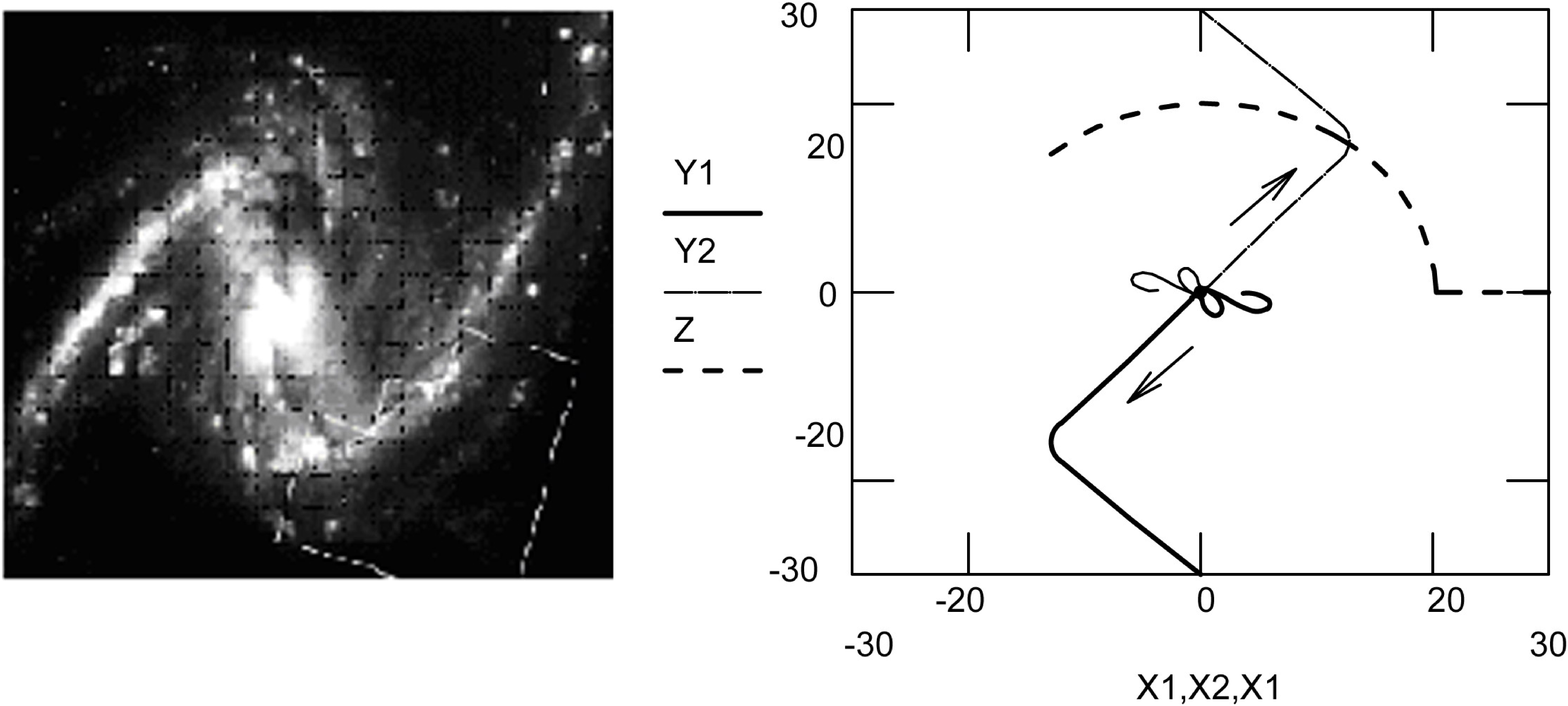,
scale=0.185}
   }
\end{center}
\vspace*{-14mm}
\end{figure}
\begin{center}
{\bf Fig.\,1\,-left\,:} \,Galaxy NGC-1365\, (Hubble telescope image, NASA/ESA)\\[1.5mm]
{\bf Fig.\,1\,-right\,:}\! Numerical calculation based on the ``center plus current''\\
model\, in the framework of \,AGD\, (the exact view\, of the central details\\ depends on the step of the calculation but they remain always present)
\end{center}
\noindent Besides, there are also the images received
recently by the space observatory HERSCHEL [72] (compare Fig.\,1\,-right and
Fig.\,2) when photographing the center of our galaxy. Thus, there could be a new approach to the study of the origins of the arms and bars, characteristic of most spiral galaxies.

\begin{figure}[htbp]
\begin{center}
 \vspace*{-1mm}
  {
  \epsfig
{file=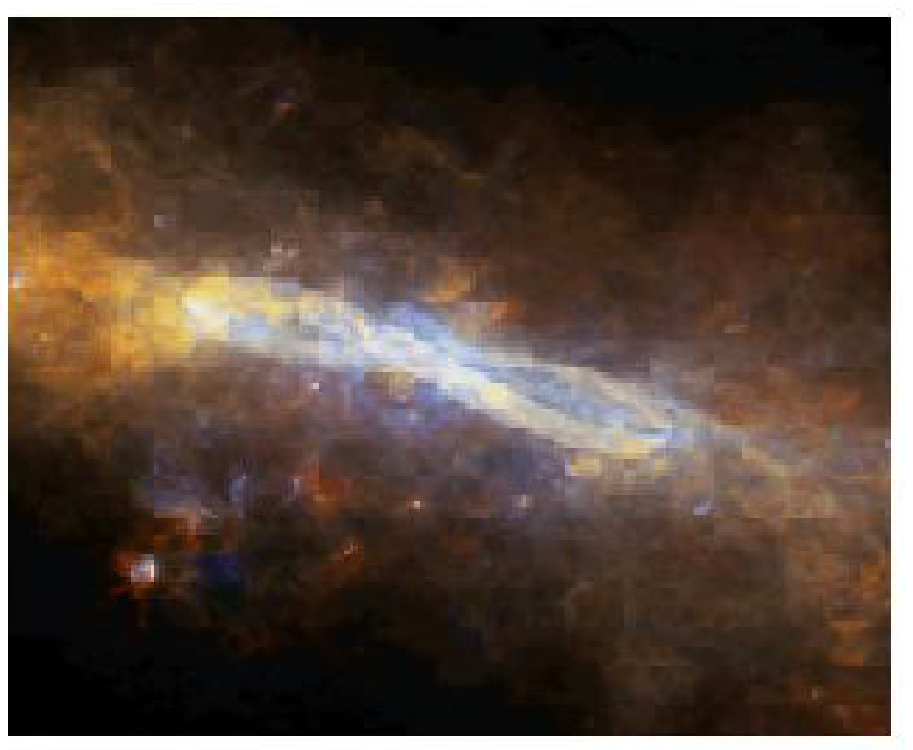,
scale=0.65}
   } 
\end{center}
\vspace*{-11mm}
\end{figure}
\begin{center}
{\bf Fig.\,2\,:} \,Details discovered by Herschel orbital observatory in \\the center of Milky Way (to appear in ApJ)
\end{center}

The theory presented in this section is based on the new notion, which serves
the basis for the description of physical reality -- the phase space-time admitting the use of various geometries to describe its subspaces. The AGD approach is consistent with observations at the galactic scale, and does not require the introduction of dark matter. Besides, it includes a new (or additional) interpretation of the Hubble law, which takes into account not the radial expansion of the Universe but the various tangential motions of its distant parts. The last being regarded in the framework of AGD leads to a linear decrease of frequency with distance to the radiation sources explained by the {\it gravitational} red shift. The observations of extremely high tangent velocities of the distant quasars present an indirect evidence supporting this idea.

\section{Mathematical, physical and geometric aspects\\ of hyper-complex numbers algebra}

The natural basis for Finsler geometries of special type (the so-called Berwald-Moor spaces
$\mathcal{H}_n$  with metric
\begin{equation}\label{26}
{}^nG=\hat{\mathcal{S}}(dx^1\otimes dx^2\dots\otimes dx^n),
\end{equation}
where $\hat{\mathcal{S}}$ is the symmetrization operator (without the numerical factor)\,) represent the well-known associative-commutative algebras $P_n.$ 

This section of the review is devoted to presenting the geometrical, algebraic and physical results obtained in the study of poly-numbers associative-commutative algebras and Berwald-Moor geometries of various dimensions related to them.

\subsection{Conformal gauges and non-linear symmetries}

It is well known that the Finslerian Berwald-Moor space $\mathcal{H}_n$ have a rich (infinite) group of conformal symmetries $\mathcal{C}\mathcal{H}_n.$ We denote by $\mathcal{H}_n^f$ the Berwald-Moor manifold in a {\it special conformal gauge},  which can be obtained from $\mathcal{H}_n$ by the action of some $f\in \mathcal{C}\mathcal{H}_n.$                                                                                                                                                                                                                                                                                                                                                                                                                                                                                                                                                                                                                                                                                                                                                                                                                                                                                                                                                                                                                                                                                                                                                                                                                                                                                                                                                                                                                                                                                                                                                                                                                                                                                                                                                                                                                                                                                                                                                                                                                                                                                                                                                                                                                                                                                                                                                                                                                                                                                                                                                                                                                                                                                                                                                                                                                                                                                                                                                                                                                                                                                                                                                                                                                                                                                                                                                                                                                                                                                                                                                                                                                                                                                                                                                                                                                                                                                                                                                                                                                                                                                                                                                                                                                                                                                                                                                                                                                                                                                                                                                                                                                                                                                                                                                                                                                                                                                                                                                                                                                                                                                                                                                                                                                                                                                                                                                                                                                                                                                                                                                                                                                                                                                                                                                                                                                                                                                                                                                                                                                                                                                                                                                                                                                                                                                                                                                                                                                                                                                                                                                                                                                                                                                                                                                                                                                                                                                                                                                                                                                                                                                                                                                                                                                                                                                                                    Instead of the transformations of the manifold $\mathcal{H}_n$ belonging to the group $\text{Iso}\mathcal{H}_n,$ we now have the transformations of the manifold $\mathcal{H}_n^f$ belonging to the group $(\text{Iso}\mathcal{H}_n)^f<\mathcal{C}\mathcal{H}_n^f,$ whose elements $\iota^f$ are defined by the formula $\iota^f\equiv f\circ\iota\circ
 f^{-1}.$ The action of the group $(\text{Iso}\mathcal{H}_n)^f$ in the coordinate space of the manifold $\mathcal{H}_n^f$ in general case is described by nonlinear functions, so this group is naturally called {\it the nonlinear $f$-representation of the group} $\text{Iso}\mathcal{H}_n.$ In general, the group $(\text{Iso}\mathcal{H}_4)^f$ can always be regarded as a (generally nonlinear) group of isometries $\text{Iso}{\mathcal{H}_4'}^f$ of a manifold
${\mathcal{H}_4'}^f$, which differs from $\mathcal{H}_4^f$, only by its metric. The form of this metric depends on the type of
the gauge function $f$. 

In [39,\,40] there are concrete examples that illustrate the fact that the isometry group and the group of conformal symmetries of the Berwald-Moor metric can interact with each other in a non-trivial way leading to nonlinear symmetries of the known geometries.

\subsection{Osculating Riemannian metrics}

With the disposal of the metric (26) and vector fields of Lie algebras of the groups $\text{Iso}\mathcal{H}_n$ (and $\mathcal{CH}_n),$ one can naturally obtain an infinite number of Riemannian metrics out of the metric (26) with the help of the following general technique. Consider the ``incomplete''\, scalar poly-product of the form\,:
 \[
 g={}^nG(X_{(1)},X_{(2)},\dots,X_{(n-2)},\ ,\ ),
 \]
 where $X_{(j)}$ are the elements of Lie algebras of the groups $\text{Iso}\mathcal{H}_n$ and (or) $\mathcal{CH}_n,$ that are for convenience numbered by the indices corresponding to their places as the arguments of the Berwald-Moor metric. It is obvious that $g$ is a (pseudo-) Riemannian metric, depending on the chosen fields $X_{(j)}.$ The described method leads to a generalization of the concept of ``Riemannian metric osculating to a given Finslerian metric''\,, discussed in [1].

 Consider as a reference vector field a common element of the Lie algebra of the subgroup
of the uni-modular dilations $\text{Iso}_D\mathcal{H}_3$ of the complete group $\text{Iso}\mathcal{H}_3,$ which has the form\,:
 \[
 X=b_1D_1+b_2D_2=b_1x^1\partial_1+(b_2-b_1)x^2\partial_2-b_2x^3\partial_3,
 \]
 where $b_1,b_2$ are arbitrary real parameters. The Riemannian metric osculating along this field
has the form\,:
 \begin{equation}
 \label{27}
 \begin{matrix}
 g=b_1x^1(dx^2\otimes dx^3+dx^3\otimes dx^2)+\\[5pt]
 +(b_2-b_1)x^2(dx^1\otimes dx^3+dx^3\otimes
 dx^1)-\\[5pt]
 -b_2x^3(dx^1\otimes dx^2+dx^2\otimes
 dx^1).
 \end{matrix}
 \end{equation}
  This metric is generally not flat. Its determinant defining a local volume element is given by\,:
 \[
\text{det}(g)=2b_1b_2(b_2-b_1)x^1x^2x^3.
 \]
 One can see that the metric (27) is nonsingular only if all of the conditions\,: $b_1\neq0,$ $b_2\neq0,$ $b_1\neq b_2\,$ are fulfilled simultaneously. The standard study of isometries and conformal symmetries
of this metric reveals the fact that this metric has a 3-dimensional algebra of isometries and
10-dimensional algebra of conformal symmetries. Such a rich algebra of conformal symmetries is a residual ``track''\, of the infinite-dimensional algebra of conformal symmetries of the original Berwald-Moor metric (26) for $n=3.$

The study of residual symmetries of the Riemannian metrics osculating to the Berwald-Moor metrics admits a more general setting in which we obtain the following basic relations\,:
\begin{equation}\label{28}
 L_{X_{i}}g_{(j)}=L_{X_{(i)}}G(X_{(j)},\ ,\ )=G([X_{(i)},X_{(j)}],\ ,\
 )=c^k_{ij}g_{(j)}
 \end{equation}
 and
 \begin{equation}\label{29}
 L_{\bar X_{i}}\bar g_{(j)}=L_{\bar X_{(i)}}G(\bar X_{(j)},\ ,\ )=\phi_i\bar g_{(j)}+\bar c^k_{ij}\bar g_{(k)}=(\phi_i\delta^k_j+\bar
 c^k_{ij})\bar g_{(k)},
 \end{equation}
 where $X_{(j)}$ is an element of the Lie algebra of the group 
$\text{Iso}\mathcal{H}_3,$ $\bar{X}_{(j)}$ is an element of the Lie algebra of the conformal group $\mathcal{CH}_3,$ $c^k_{ij}$ and $\bar c^k_{ij}$ are the structural constants or the structural functions of the Lie algebra of the groups $\text{Iso}\mathcal{H}_3$ and $\mathcal{CH}_3$ respectively.

Thus, {\it the families of metrics $\{g_{(j)}\}$ and $\{\bar g_{(j)}\}$ form differential ideals with respect to their Lie differentiation along the families of fields $\{X_{(j)}\}$ and $\{\bar X_{(j)}\}$ respectively.} This is a general property and it allows us to formulate some general theorems concerning the symmetry of Riemannian metrics $g_{(j)}$ and $\bar g_{(j)}$ [41].

\subsection{Metric bingles in $\mathcal{H}_3$} 

Studying the properties of angles in Finsler geometry is of particular interest for its physical applications. One of the approaches to the problem of constructing of additive poly-angles (e.g. bingles and tringles) is to formulate and solve the corresponding functional equations that satisfy the additivity condition [42]. Instead of solving the functional differential equations in the space of basic conformal invariants of B-M geometry, one can from the very beginning relate all the types of poly-angles with notions additive by their definition, such as lengths, areas or volumes, calculated on the unit sphere (indicatrix) of the B-M geometry.

It turns out that for any pair of non-isotropic vectors $A$ and $B$ one can introduce two types of bingles -- mutual and relative. The expression for the mutual bingle has the following form\,:
\begin{equation}\label{30}
\phi[A,B]=|A^{\flat}-B^{\flat}|
\end{equation}
where $\flat$ is a bi-projection operation in $\mathcal{H}_3,$ which acts on an arbitrary element $X\in\mathcal{H}_3$ according to the rule\,:
\[
(X^\flat)^i=\ln\frac{X^i}{|X|}\,.
\]
The norm in (30) is calculated with the help of the standard Berwald-Moor metric in the isotropic coordinates. The bingl defined by (30) is additive by definition, i.e. for any triplet of the  ``coplanar'' vectors $A,B,C$ there is a condition which is analogous to the Euclidean one\,:
\begin{equation}\label{31}
\phi[A,C]=\phi[A,B]+\phi[B,C]\,.
\end{equation}
The condition of coplanarity of the vectors $A,B,C$ has the form of the condition of collinearity
of the corresponding $\flat$-images\,:
\begin{equation}\label{32}
(A^{\flat}-B^{\flat})\wedge(A^{\flat}-C^{\flat})=0.
\end{equation}
Expressions for the second (mutual) bingle (there may be three types of it, depending on the
mutual orientation of vectors $A$ and $B$) have the form\,:
\begin{equation}\label{33}
\text{cfh}\,\psi_i[A,B]=e^{(B^{\flat\flat})_i-(A^{\flat\flat})_i},\ \ i=1,2,3,
\end{equation}
and the function, which is inverse to the Finsler-hyperbolic cosine $\text{cfh},$ is defined by the integral\,:
\begin{equation}\label{34}
\text{arccfh}(\xi)\equiv-\frac{1}{2}\int\limits_{2^{-1/3}}^{\xi}\!\!
\left(\frac{(x^2-\sqrt{x(x^3+4)})(3x^2+\sqrt{x(x^3+4)})(\sqrt{x(x^3+4)}+x^3-2)}{x^4(x^3+4)}\right)^{\!1/3}\!\!dx.
\end{equation}
Finally, the expression for the value analogous to the solid angle on the vectors $A,B$ and $C$ is given by the following integral\,:
\begin{equation}\label{35}
\Sigma(A,B,C)=\frac{3}{2}\phi^2[A,B](\text{cfh}\,\psi_1[B,C]\text{cfh}\,\psi_1[A,B]-\text{cfh}\,\psi_2[B,C]\text{cfh}\,\psi_2[A,B])^2\times
\end{equation}
\[
\int\limits_{\text{cfh}\psi_1[A,B]}^{\text{cfh}\psi_1[A,C]}\frac{dx}{\sqrt{x^4+4x}\left(x/\text{cfh}\,\psi_1[B,C]-(\sqrt{x^4+4x}-x^2)/(2x\text{cfh}\,\psi_2[B,C])\right)^2}\,.
\]
The exact formulations, proofs and illustrations can be found in [43].

\subsection{Classification of homogeneous cubic metrics}

Symmetry analysis of geometric objects is a key means of study of their internal invariant properties (i.e. being independent on the coordinates). In order to understand the place of Berwald-Moor metric among other related cubic metrics, the study of the isometry group of the general homogeneous cubic form
\begin{equation}\label{36}
G=G_{\alpha\beta\gamma}dx^\alpha\otimes dx^\beta\otimes dx^\gamma,
\end{equation}
was undertaken. Here $G_{\alpha\beta\gamma}$ are the constant real components of the cubic form. The results of the study are summarized in the following table\,:

\medskip

\begin{center}
\begin{tabular}{|c|c|c|c|c|c|c|c|c|}
\hline
Symmetry classes& 1&2&3&4&5&6&7&8\\
\hline
Projective classes&III,XII&V&(1): VIII, (2): & IV&
II,X,XI&?&---&Gen., I,IX\\
&&&VI,XIII, (3): VII&&&&&\\
\hline
\end{tabular}
\end{center}

\medskip

This proves that the symmetry analysis reveals 6 different symmetry classes (7th class is empty, and the 6th coincides with the 5th), the previously known 13 projective classes [44] are distributed amongst them. The Berwald-Moor metric falls into the 1st symmetry class. One of the important findings of this study is the conclusion on the incompleteness of the classification of cubic homogeneous metrics according to their isometry algebras [45].

\subsection{$h$-holomorphic functions of a double variable}

For the interpretation of ${\mathbb{R}}^2$ as a plane of a double variable $\mathcal{H},$ it is natural to consider only the maps that preserve the hyperbolic complex structure of the plane, i.e. by the maps $\mathcal{H}\to \mathcal{H}$ of the form: $h\mapsto
s=F(h).$ The differentiable functions ${\mathbb{R}}^2\to {\mathbb{R}}^2$ which satisfy the condition\,: $F_{,\bar h}=0,$ are called $h$-holomorphic functions of double variable $h.$\\
Let us formulate some important properties of the $h$-holomorphic functions as theorems.
\medskip
\\
\textbf{Theorem 1.} {\it Any $h$-holomorphic function maps zero divisors into zero divisors.}
\medskip
\\
{\bf Theorem 2.} {\it The components $U$\! and $V$\! of the $h$-holomorphic function $F=U+jV$ satisfy
the hyperbolic Cauchy-Riemann conditions: $U_{,t}=V_{,x}\,;\quad U_{,x}=V_{,t}\,.$}
\medskip
\\
{\bf Theorem 3.} {\it For any $h$-holomorphic function $F$ in $D,$ there holds the integral Cauchy Theorem\,:
\[
\oint\limits_{\Gamma}F(h)\, dh=0,
\]
where $\Gamma$ is a simple closed piecewise smooth contour which has no isotropic elements and lies entirely in $D.$}
\medskip
\\
{\bf Theorem 4.} {\it For any $h$-holomorphic function $F$ in $D,$ there holds true the integral Cauchy formula\,:
\[
\oint\limits_{\Gamma}\frac{F(h)}{h-h_0}\, dh=0,
\]
where $\Gamma$ is a simple closed piecewise smooth contour which has no isotropic elements, lies entirely in $D$ and encloses the point $h_0.$}\\
Other versions of Cauchy's integral formula are given in [46].
\medskip
\\
{\bf Theorem 5.} {\it For a simple closed piecewise smooth contour $\Gamma$ that has no isotropic elements and encloses the point $h_0,$ we have the formula:
\begin{equation}\label{37}
\oint\limits_{\Gamma}(h-h_0)^\alpha\, dh= \left\{
\begin{array}{lr}
0,&\alpha\neq-1;\\
j\ell_H,&\alpha=-1.
\end{array}
\right.\end{equation}
where $\alpha$ is any real number, $\ell_H$ is the improper ``fundamental constant''\, in the plane of double variable that determines the amount of space of the hyperbolic angles (analogous to the constant $2\pi$ in the complex plane).}
\medskip
\\
{\bf Theorem 6.} {\it The pseudo-Euclidean metric $\eta=\text{Re}(dh\otimes d\bar h)$ is conformal relative to an arbitrary $h$-holomorphic mapping of the plane of the double variable.}
\medskip
\\
In [46] the properties of the basic elementary $h$-holomorphic functions of the double variable were studied in detail.

\subsection{Hyperbolic field theory on the $\mathcal{H}_2$ plane}

We consider an arbitrary $h$-holomorphic function $F(h)=U+jV$ as complex $h$-potential of a certain 2-dimensional vector field ({\it $h$-field}) in the plane of double variable. The real part $U$ of this function we associate with the potential of the field ({\it $h$-potential function}) and the imaginary part $V$ we associate with {\it the strength function} of this field. We define the strength, $\mathcal{E}$ of the $h$-field by the formula\,:
\begin{equation}\label{38}
\mathcal{E}=\mathcal{E}_t+j\mathcal{E}_x=-\overline{\frac{dF}{dh}}=-\frac{d\bar
F}{d\bar h}=-U_{,t}+jU_{,x}\,,
\end{equation}
which can be regarded as a double form of representation for the vector field of the gradient of the function $U$ with respect to the pseudo-Euclidean metric. Equation (38) is obtained taking into account the hyperbolic Cauchy-Riemann conditions.

In view of the relation $\mathcal{E}=\mathcal{E}(\bar z)$ (antiholomorphicity of strength), arising from the definition (38), we obtain the following identity\,:
\begin{equation}\label{39}
\frac{\partial \mathcal{E}}{\partial
h}=\frac{1}{2}[\mathcal{E}_{t,t}+\mathcal{E}_{x,x}+j(\mathcal{E}_{t,x}+\mathcal{E}_{x,t})]=0,
\end{equation}
which is equivalent to two identities\,:
\begin{equation}\label{40}
\text{divh}\, \mathcal{E}\equiv
\mathcal{E}_{t,t}+\mathcal{E}_{x,x}=0;\quad \text{roth}\,
\mathcal{E}\equiv \mathcal{E}_{t,x}+\mathcal{E}_{x,t}=0,
\end{equation}
expressing, respectively, {\it the solenoidal} and {\it $h$-potential} properties of the electrostatic field\footnote{Notice that the divergence of the vector field is defined in the same way in the complex and hyperbolic cases, as opposed to the operation of the curl of a vector field, which includes the symmetric combination of partial derivatives in the hyperbolic case.}.

As an example, consider the $h$-potential of the form
\begin{equation}\label{41}
F(h)=-q\ln h\,,
\end{equation}
which is obviously the hyperbolic generalization of the Coulomb potential. The corresponding field strength is given by (38) and has the form\,:
\begin{equation}\label{42}
\mathcal{E}=\frac{q}{\bar
h}=\frac{qh}{|h|^2}=q\left(\frac{t}{t^2-x^2}+j\frac{x}{t^2-x^2}\right).
\end{equation}
 The field lines of a hyperbolic point source are the radial lines with $\psi=\text{const},$ and the equipotential lines are the hyperbolas $\varrho=\text{const}.$ The picture of the field lines in all 4 wedges is shown on the Figure\,:
\begin{figure}[h]
\centering
        \input{hq1.pic}
\end{figure}

The dual interpretation of the hyperbolic point source is obtained by passing from the potential $F(h)$ in (41) to the potential $jF(h).$ At the same time for a new dual field $\mathcal{B}$ we 
get the following expression\,:
\newpage
\vspace*{-1cm}
\begin{equation}\label{43}
\mathcal{B}=j\frac{d\bar F}{d\bar h}=-\frac{qj}{\bar
h}=-q\frac{x+jt}{t^2-x^2}.
\end{equation}
Field $\mathcal{B}$ is a hyperbolic analogue of a point vortex. Its lines of force are shown in the picture and present the hyperbolas\,:
\begin{figure}[h]
\centering
        \input{hc2.pic}
\end{figure}\\
By analogy with the complex case, it is possible to combine the above two situations into
one introducing the concept of {\it hyperbolic vortex-source} with the complex charge $\mathcal{Q}=q-jm.$ Then the potential takes the form:
\vspace*{-2mm}
\begin{equation}\label{44}
F(z)=-\mathcal{Q}\ln h=-q\ln\varrho+m\psi-j(-m\ln\varrho+q\psi).
\end{equation}

\vspace*{-2mm}
\noindent Such a potential can be most naturally interpreted in the framework of dual-symmetric
hyperbolic field theory in which the hyperbolic electric and magnetic charges and currents are present on ``equal footing''\,. The equation for the field lines of such field is obtained from (44) by equating the imaginary part to a constant\,:
\vspace*{-2mm}
\begin{equation}\label{45}
(t+x)^{1-\alpha}(t-x)^{1+\alpha}=\text{const},
\end{equation}

\vspace*{-3mm}
\noindent where $\alpha=q/m.$ The picture of the field lines for $\alpha=-2$ is shown on the Figure\,:\\

\begin{figure}[htbp]
\begin{center}
\vspace*{-5mm}
  {
  \epsfig{file=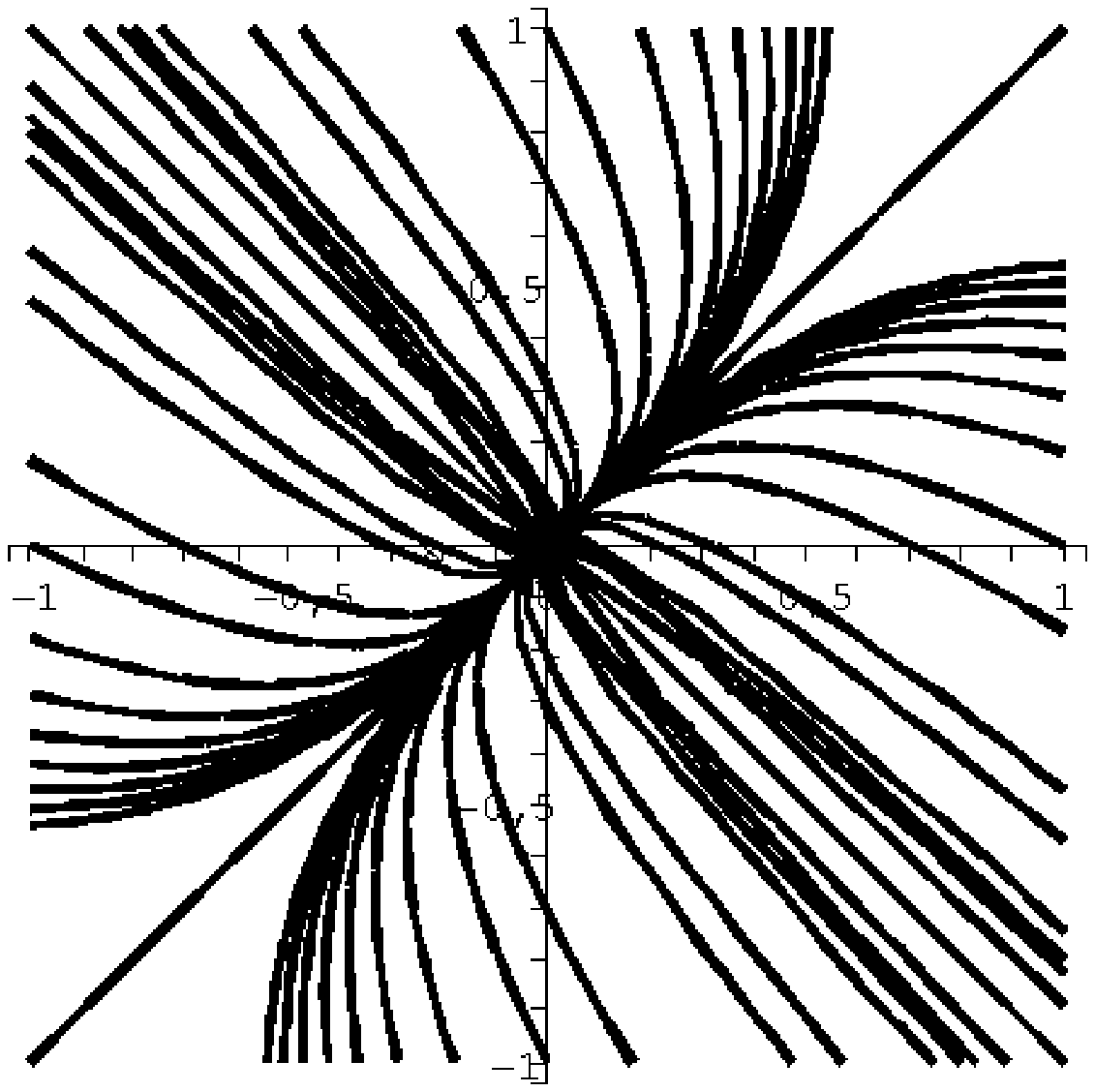, scale=0.65}
   } 
         \end{center}
\end{figure}
\vspace*{-5mm}
For physical applications it is necessary to generalize the concept of the $h$-field for the case of commutative and associative algebras of higher dimensions. In what follows we illustrate the idea of such a generalization by the example of the algebra of 3-numbers $P_3.$
\newpage
We start with an isotropic basis in the $P_3,$ in which the $h$-holomorphic function has the following representation\,:
\begin{equation}\label{46}
F(h)=F(\xi_1)e_1+F(\xi_2)e_2+F(\xi_3)e_3.
\end{equation}
The operators of differentiation with respect to the independent variables $h,h^\dag,h^\ddag$ have the following form\,:
\begin{equation}\label{47}
\frac{\partial}{\partial
h}=e_1\frac{\partial}{\partial\xi_1}+e_2\frac{\partial}{\partial\xi_2}+e_3\frac{\partial}{\partial\xi_3};\
\frac{\partial}{\partial
h^\dag}=e_1\frac{\partial}{\partial\xi_3}+e_2\frac{\partial}{\partial\xi_1}+e_3\frac{\partial}{\partial\xi_2};\
\frac{\partial}{\partial
h^\ddag}=e_1\frac{\partial}{\partial\xi_2}+e_2\frac{\partial}{\partial\xi_3}+e_3\frac{\partial}{\partial\xi_1}.
\end{equation}
With the help of (47), one can easily verify the validity of equations for function $F$ in the form of (46) by the direct calculation in components
\begin{equation}\label{48}
\frac{\partial F}{\partial h^\dag}=\frac{\partial F}{\partial
h^\ddag}=0;\quad \frac{\partial F}{\partial h}=\frac{\partial
F_1}{\partial \xi_1}e_1+\frac{\partial F_2}{\partial
\xi_2}e_2+\frac{\partial F_3}{\partial \xi_3}e_3,
\end{equation}
where here and further $F_i\equiv F(\xi_i)$ is the same function of various isotropic variables.
The conditions of holomorphicity (multidimensional analogue of the standard Cauchy-Riemann conditions), in symmetric non-isotropic basis $\{j_1,j_2,j_3\}$  which is defined by\,:
\begin{equation}\label{49}
j_1=e_1-e_2-e_3;\quad j_2=-e_1+e_2-e_3;\quad j_3=-e_1-e_2+e_3
\end{equation}
and by the rules of multiplication\,:
\begin{equation}\label{50}
j_i^2=-(j_1+j_2+j_3);\quad j_i\cdot j_k=j_l\quad (j\neq k\neq l)\,,
\end{equation}
have the form of matrix differential equations\,:
\begin{equation}\label{51}
\left(\begin{array}{ccc}
-(\bar\partial+\partial_3)&\partial_{2-1}&\partial_{1-2}\\
\partial_{2-3}&-(\bar\partial+\partial_1)& \partial_{3-2}\\
\partial_{1-3}&\partial_{3-1}&-(\bar\partial+\partial_2)
\end{array}\right)\,
\left(\begin{array}{c}
U_1\\
U_2\\
U_3
\end{array}\right)=0,\quad
\end{equation}
\begin{equation}\label{52}
\left(\begin{array}{ccc}
-(\bar\partial+\partial_2)&\partial_{1-3}&\partial_{3-1}\\
\partial_{1-2}&-(\bar\partial+\partial_3)& \partial_{2-1}\\
\partial_{3-2}&\partial_{2-3}&-(\bar\partial+\partial_1)
\end{array}\right)\,
\left(\begin{array}{c}
U_1\\
U_2\\
U_3
\end{array}\right)=0
\end{equation}
for every $h$-holomorphic function $F(h)=U_1j_1+U_2j_2+U_3j_3.$ Here $\partial_{i-j}\equiv\partial_i-\partial_j,$
$\bar\partial=\partial_1+\partial_2+\partial_3.$ Due to the invariance properties of the $h$-holomorphy with respect to the choice of the algebra basis, we can say that the general solution of (51)-(52) is written by representing $U_i$ in terms of $F_i$ (components in the isotropic basis) expressed in terms of $x$-coordinates\,:
\[
U_1=F(x_2-x_1-x_3)+F(x_3-x_1-x_2);\quad
U_2=F(x_1-x_2-x_3)+F(x_3-x_1-x_2);\quad
\]
\begin{equation}\label{53}
U_3=F(x_1-x_2-x_3)+F(x_2-x_1-x_3).
\end{equation}
This fact can be verified by direct substitution of (53) into (51)-(52). The combinations of coordinates in the arguments of $F$ present the higher analogues of retarded and advanced arguments in the double plane.

The third-order operator
\begin{equation}\label{54}
\Delta^{(3)}\equiv\frac{\partial}{\partial
h}\,\frac{\partial}{\partial h^\dag}\,\frac{\partial}{\partial
h^\ddag}=(e_1+e_2+e_3)\frac{\partial}{\partial
\xi_1}\frac{\partial}{\partial \xi_2}\frac{\partial}{\partial
\xi_3}
\end{equation}
is proportional to the algebraic unity, so for every smooth function $F(h,h^\dag,h^{\ddag})$
\begin{equation}\label{55}
\Delta^{(3)}F=(\Delta^{(3)}U_1)j_1+(\Delta^{(3)}U_2)j_2+(\Delta^{(3)}U_3)j_3\,.
\end{equation}
If the function $F$ is $h$-holomorphic, then because the operator $\Delta^{(3)}$ contains differentiation with respect to $h^\dag$ and $h^\ddag,$ there holds the relation $\Delta^{(3)}F\equiv0\,,$ which is equivalent to its three components\,:
\begin{equation}\label{56}
\Delta^{(3)}U_i\equiv0\quad (i=1,2,3)\,.
\end{equation}
Equation (56) is a 3-dimensional analogue of the harmonicity conditions or the hyperbolic
harmonicity conditions which are identically satisfied by holomorphic functions of complex or of double variable, respectively.

The discussion and development of these ideas can be found in [47].

\subsection{Conformal two-dimensional theory of relativity}

We extend the Poincare group acting on the two-dimensional space-time $\mathcal{M}_2$ to a group of arbitrary $h$-holomorphic transformations that operate on points-events of space-time as on the elements of $\mathcal{H}_2.$ Using the exponential representation for the derivative of $F'$\,:
\begin{equation}\label{57}
F'(h)=\epsilon|F'|(t,x)e^{j\psi(t,x)},
\end{equation}
we conclude that locally $h-$holomorphic transformations implement not only reflections and boosts known in the theory of relativity but also the extension of lengths of the vectors (scalar factor  $|F'|(t,x)$). Let us consider the function $F=U+jV$ as the complex potential of the reference vector field of the 2-velocity or {\it the reference field of the proper time}. The field of 2-velocity $u$ is determined by the formula\,:
\begin{equation}\label{58}
u={\frac{dF}{dh}}=\frac{\partial U}{\partial t}+j\frac{\partial
U}{\partial x},
\end{equation}
which uses the definition of the operator of complex differentiation and the hyperbolic Cauchy-Riemann conditions. The square of the modulus of the 2-velocity is
\begin{equation}\label{59}
|u|^2=(\triangledown U)^2=(\triangledown V)^2=|F'|^2.
\end{equation}
``The velocity field''\, of the proper time for any integral curve $\Gamma$ of this field is given by\,:
\begin{equation}\label{60}
\frac{d\tau}{ds}=|F'|.
\end{equation}
Now in the $h$-holomorphic theory of relativity under consideration, the intervals of the pseudo-Euclidean length and time become different and the relationship between them at each point is governed by the hyper-complex potential $F.$

The integral curves of the field  $\triangledown V$ are  spatial sections of 2-dimensional space-time, orthogonal to the lines of time at every point. Thus, the scale factor ``governs''\, both the course of the proper time and spatial distances.

For arbitrary motions of test particles, the length and time intervals are calculated as\,: \begin{equation}\label{61}
\frac{d\tau}{ds}=\eta(\triangledown U,w);\quad
\frac{d\ell}{ds}=\eta(\triangledown V,w),
\end{equation}
where $w$ is the standard vector of the 2-velocity of the test particle $(|w|=1).$

The simplest version of the variational principle of the dynamics theory of the hyperbolic field that takes into account the non-holomorphy of the hyperbolic potential inside the sources is determined by the action of the form\,:
\begin{equation}\label{62}
\mathcal{S}[F,\bar
F]=\alpha\int\limits_{\mathcal{H}_2}\{\left|F_{,h}\right|^2-\mathcal{U}(\left|F_{,\bar
h}\right|^2)\}\, dh\wedge d\bar h,
\end{equation}
where the first term under the integral is a hyperbolic ``kinetic term''\,. It is responsible for the dynamics of the hyperbolic potential in vacuum. The second term represents a hyperbolic ``potential term''\, and is responsible for the properties and for the contribution of sources. This last term depends only on the hyperbolic modulus of the magnitude of non-holomorphy, and in the region outside the sources, where the non-holomorphy becomes equal to zero, it defines (in the action) a certain ``full divergence'' that does not give any contribution to the equations of motion. The standard procedure of varying the action (62) over the field variables $\,\bar F,\,F\,$ leads to the following field equations\,:
\begin{equation}\label{63}
\frac{1}{4}\square F=(\mathcal{U}'F_{,\bar h})_{,h}\,.
\end{equation}
This expression is the inhomogeneous wave equation with a source on the right-hand side, depending only on the non-holomorphy of $F.$ As expected, the field equations are nonlinear, since the field $F,$ as follows from the principles of the theory, describes its own sources through effective self-interaction. In this sense the developed theory is adjacent to the versions of the unified field theory by Mie.

A remarkable feature of equations (63) is the existence (regardless of the specific form of the potential function $\mathcal{U}$) of the first integral 
\begin{equation}\label{64}
F_{,\bar h}(1-\mathcal{U}')=\varphi(\bar h)
\end{equation}
containing an arbitrary function $\varphi(\bar h).$

The explicit expressions for the energy density of the algebraidized matter $\varepsilon$ and its pressure $p,$ obtained using the standard formalism of the field theory (Noether's theorem), have the form\,:
\begin{equation}\label{65}
\varepsilon=\mathcal{U}-\mathcal{U}'X+(1-\mathcal{U}')\sqrt{XY}\,;\quad
p=\mathcal{U}-\mathcal{U}'X-(1-\mathcal{U}')\sqrt{XY}\,,
\end{equation}
where $Y\equiv |F_{,h}|^2.$

With the help of the super-variational principle introduced in [48], it appears possible to calculate the general form of the potential $\mathcal{U}$ in the theory presented here
\begin{equation}\label{66}
\mathcal{U}(X)=3X+{U}_0+2U_1\ln\left|1-\frac{X}{U_1}\right|\,,
\end{equation}
where $U_0$ and $U_1$ are two fundamental constants of the theory.

\section{Differential-geometric aspects of the theory of\\ Berwald-Moor type Finsler spaces of\\ various dimensions}

	In order to find out the fundamental relations between the scalar poly-product and geometric objects induced by it, as          a continuation of research on these interrelations (which started by the works of M.\,Matsumoto, H.\,Shimada, S.\,Numata, K.\,Okubo and of Romanian geometers [56, refs. no. [29-32] and 43] and [61/refs.[3-4]]), new correlations were obtained between the Berwald-Moor $m$-th root pseudo-norms and geometric objects from the classical Finslerian context ([61, \S 2 and 5-7]). Such relations were investigated in [60] and [58]. Their role is a methodological one: they enhance the process of deriving properties of certain structures (e.g., projective ones, [61]), or passing from algebraic aspects of $m$-th root metric theory to differential geometry specific aspects from the theory of Finsler spaces.

\subsection{Description of the obtained results}

	The study of connections which are compatible with remarkable geometric structures was performed in [58], where, for specific connections from Finsler geometry (e.g., for Cartan, Barthel and Miron connections), the authors point out the properties of induced connections on hypersurfaces, as a necessary step in the study of mean $Y$-curvature within the $N$-extremality framework. In this study, the authors propose an original software for the calculation and the use of Finslerian geometric objects specific to the study of $y$-minimal submanifolds. With the help of Maple symbolic calculations, they determine the coefficients of these geometric objects for low-dimensional manifolds equipped with 3-rd and 4-th root metrics and with Berwald-Moor conformal metrics. The underlying algorithm of this Maple software, was introduced by M.\,Matsumoto ([58/ refs. [39-40]]) and is likely to provide a wide range of applications in the study of anisotropic media. 
	
	Moreover, in the paper [61, \S 3], there are indicated the essential connections to be used in determining whether an $m$-th root metric space is of Berwald or of Douglas type, and are derived original results concerning spaces with Berwald-Moor type metrics.
	
	On the other side, in [54], Landsberg spaces are characterized by means of classical connections (Vasiman and Levi-Civita) and the connections which are subsequently induced into the structural and transversal fibers of the bundle, are indicated.
	 
	The work [49, \S 2-3] performs a preliminary study of the cotangent bundle of spaces endowed with Berwald-Moor metric; in the cited paper, one defines the $v$-curvature tensors and the $T$-tensor of the Berwald-Moor space, for the case of Shimada-type $m$-th root metrics; the results are specialized for the case of dual $m$-th root metric spaces having the indicatrix given by the product of the momentum components. In this case, the classical results regarding the vanishing of the torsion tensor and of the $T$-tensor, obtained by M.\,Matsumoto and H.\,Shimada ([49/ refs. [5,\,11]]) and also the property of $S3$-likeness for the Berwald-Moor space, were obtained for the first time for the dual case.
	 
	The determining of connections and of induced geometric objects on submanifolds of $m$-th root metric spaces was carried out in the paper [58], by summing up known results and by original implementation of their construction into Maple code, using macros and supplementary procedures which simplify the use of the code and allows to extend the results specific to the study of the indicatrix.
	
	The procedure for obtaining the mean curvature and minimal ($Y$-extremal) surfaces/hypersurfaces and corresponding computer simulation are presented in [58], which embeds two addenda devoted to the 4-dimensional case. Here, the mean curvature and the equations of $Y$-extremal (hyper-)surfaces are explicitly obtained by the use of symbolic software, and the calculation of the explicit form of the normal field to a submanifold (theoretically described in [58/ ref. [40]]), represents a concrete application of software procedures in solving nonlinear equations. Also, the mean curvature -- depending on the energy of a space-like or light-like normal vector field, is obtained by using specific procedures of the relativistic pseudo-Finslerian approach. This approach imposes restrictions on the submanifolds for the indicated practical applications, aiming to find solutions of the equations of $Y$-extremal submanifolds.
	 
	Another aim of an earlier planned research ([54/ ref. [2]]) relates to determining specific types of cohomology in $m$-th root metric spaces ([63,54]), including the cohomologies of certain Berwald-Moor Finslerian spaces. These papers present new results concerning fibered structures of Finsler type: in [63], it is proved the existence of a diffeomorphism between the 2-jet vertical bundle induced by the canonical bundle and the product of the horizontal, vertical and mixed subbundles of 2-jets induced by the second order tangent bundle. In [54], it is introduced the Vaisman connection and it is proven that the pair Levi-Civita connection -- Vaisman connection induces a pair of connections of the same type as the initial ones in the structural bundle, only if the base manifold $M$ is a Landsberg space, and that the slit tangent bundle (the tangent bundle without the image of its null section, denoted as $TM^0$) is a Reinhart space if and only if the base manifold is Riemannian. Further, the 2-leaf jets on $TM^0$ are studied, it is obtained a decomposition of this space, and the 1-dimenional Cech cohomology group with coefficients from the sheaf of basic functions is constructed in terms of fields on leaves of 2-jets. It is defined the Mastrogiacomo coholology group with respect to the connection on the structure sheaf induced by a connection on the manifold $TM^0$ and it is proven that the associated cohomology group is isomorphic to the 1-dimensional Cech cohomology group on the manifold $TM^0$, having as coefficients germs of functions on $TM^0$, which are related to the induced connection; in particular, for 4-dimensional $m$-th root spaces, it is proven that this sheaf is isomorphic to the sheaf of basic functions on $TM^0$.
	
	In  [65, 64],  the $HC$($n$) Halphen-Castelnuovo problem for smooth curves is split into two parts: the study of the lacunary and of the non-lacunary domain. The latter one is studied: existence obstructions are determined and examples of curves on rational surfaces and irrational scrolls are built. There are studied for the first time the existence of smooth irreducible non-degenerate curves of degree $d$  and genus $g$  from the projective space (the Halphen-Castelnuovo $HC$($n$) problem). For the domains $D_1^n$ and $D_2^n$ , built in the plane ($d; g$), it is shown that $D_1^n$ is simply connected, by using curves which are displaced on rational surfaces related to hyper-elliptic type sections, and are presented well known theorems of Ciliberto, Sernesi and P\u as\u arescu. As well, using results of Horrowitz, Ciliberto, Harris and Eisenbud, it is conjectured that $D_n$ is exactly the targeted lacunary domain.
	 
	Geodesics and Jacobi fields are investigated in [61, \S 2] and [60, \S 3], where geodesic equations and spray coefficients are introduced and studied for conformal flag metrics. In [61], the $hv$-curvature tensor is determined for arbitrary $m$-th root structures; this result is further used in determining the specific characterizations of Landsberg and Berwald-type $m$-th root metric spaces. Relations between the coefficients of two sprays for non-decomposable metrics are obtained in an explicit form, for cubic metrics in [60, \S 6]. All these results complete known results obtained for $m$-th root metric spaces by M.\,Matsumoto and H.\,Shimada. It is emphasized the role of flag curvature, which is a key one in describing the behavior of geodesics. This is continued in [59], where there are described the geodesic equations perturbed by the presence of an electromagnetic field. In Finsler spaces, the 4-potential 1-form is anisotropic, and represents a horizontal 1-form on the tangent bundle, having specific properties.
	
	Relativistic models based on 4-dimensional $m$-th root metric are subject of intensive research. In [66], original results were obtained for the OMPR ({\em optic-metrical parametric resonance}) effect, with applications to relativity theory and to experimental physics (detection of gravitational waves). This work has been carried out in collaboration with the Russian physicist S.\,Siparov. The influence of a weak deformation of a flat pseudo-Finslerian metric upon the electromagnetic field tensor is studied in [59] and, in particular -- for the case of $m$-th root metrics of Berwald-Moor type. The generalized geometric models are obtained and the physical meaning of such a generalization, together with its role in the equations of electromagnetism in Finsler spaces is pointed out. Geodesics and Jacobi fields are also studied in [56, \S 1-2] in the context of structural stability of second order differential equations, where the authors obtain original results for sheaves of curves and for the forces which deviate trajectories from geodesics in the case of conformal deformations of $m$-th root structures or locally Minkowski metrics. With the help of supplementary software designed to determine geodesics by means of the computer, original procedures for defining the invariants which characterize the stability of structures were derived.
	
	In [61, \S 3-6], there are studied $m$-th root Berwald and Landsberg spaces and projectively flat spaces and, in the work [60, \S 4] -- cubic spaces. In [61, Th.\,17 and Th.\,18], there are investigated $m$-th root projective spaces and, in particular, Riemann-projective spaces with $m$-th root metrics [61, Th.\,19,\,20 and Prop.\,22]. All these results are original and they complete, in the case of $m$-th root metrics, known results obtained by S.\,Bacso, Zs.\,Szabo, L.\,Tamassy and Cs.\,Vincze.
	
	In [56], the authors present the basic notions from the theory of structural stability (KCC -- Kosambi-Cartan-Chern) --  created and developed by P.L.\,Antonelli, I.\,Bucataru and S.\,Sabau [56/ refs. [1-8, 48-49]] and carried out by V.\,Balan and I.R.\,Nicola for biological and ecological models ([56/ refs. [10-13]]). The five KCC-invariants are described and in the Appendix, original Maple programs for determining the invariants describing Jacobi stability of dynamical systems associated with the Finslerian approach, are presented. Sections \S 4-5 contain original results for the case of conformal deformations of $m$-th root metrics and describe the properties of the associated deformation algebras.
	
	The papers [52,\,50] extend known results for symmetric positive definite tensors for $Z$, $H$ and $E$- spectra of Berwald-Moor and Chernov tensors in 4-dimensional spaces. The algebraic properties of these tensors induce geometrical properties: by spectral techniques it is shown that the indicatrices of the associated Finsler metric are not ruled and compact, and that the problem of the minimal distance between the origin of the frame and the indicatrix has a solution depending on the $Z$-eigenvalue with maximal absolute value and with the corresponding direction given by the generating $Z$-eigenvector. The problem of asymptotic behavior of the indicatrix is solved by means of spectral properties of the symmetric tensor associated to the fundamental Finsler function. There are defined: recession vectors, degenerate vectors, singular points of the indicatrix and the best first-order approximation. The qualitative description of the three types of eigenvalues is obtained in [57] with the help of the theory of resolvents for the cubic Berwald-Moor metric.
	
	Hamilton equations, Legendre duality and physical models associated with m-th root metric spaces on the tangent or on the cotangent bundle are studied in [53], where it is indicated an essential parallelism between different transformations with physical meaning, and it is emphasized a Legendre-type relation between the Lagrangian formalism and the Hamiltonian one. It is investigated the alternative given by the use of the Rashevskii transformation commonly used in mechanics and its degenerate nature. The hyperbolic character of the Finsler and Cartan functions is emphasized and the correspondence between the basic geometric objects and their duals given by Legendre-Finsler duality for general Berwald-Moor metrics of arbitrary dimension is described.
	
	In [69,\,70,\,67,\,68], the authors build models for the gravitational and for the electromagnetic fields, based on generalized Lagrange metrics and in particular, on the locally Minkowskian Berwald-Moor metric.
	
	In [51], it is studied the geometry of submanifolds in $m$-th root metric spaces and in [53] the hyperbolic character of the Berwald-Moor metric is emphasized; in [51], the author proposes a pseudo-Finslerian formalism for Finsler metrics of locally Minkowski type, including the Berwald-Moor metric; a special attention is paid to the objects which allow to characterize minimal surfaces. Linear and nonlinear Cartan connections are studied and Gauss-Weingarten, Gauss-Codazzi, Peterson-Mainardi and Ricci-Kuhn equations are obtained. In [52,\,50], geometric properties of the Berwald-Moor indicatrix are obtained by means of spectral theory associated to a supersymmetric tensor; the spectra are obtained with the help of the Maple software.
	
	The study of cohomology classes for $m$-th root metric spaces and the study of associated bundles extend known results by addressing to the initial context of Finsler spaces (in particular, for locally Minkowskian metrics). The results include an explicit description of the Vaisman connection for a vertical fiber with respect to the verical bundle and a proof of the fact that its leaves are Reinhart spaces [62]. Further, in [63], in a vertical fiber of a Finsler manifold it is defined an adapted basis to the Liouville fibration, the vertical bundle of 2-jets and the leaves of vertical bundles of 2-jets of transversal and mixed types. It is proven that there exists a canonical diffeomorphism between the total space of the vertical bundle of 2-jets and one of the product bundles of vertical, transversal and mixed leaves of 2-jets.
	
	Specific variational features of the energy in $m$-th root metric spaces and the behavior of geodesics is studied in [69,\,70]; the authors investigate extensions of Lorentz geodesics to generalized Lagrange relativistic models obtained for small deformations of Berwald-Moor and locally Minkowski metrics. In [70], it is described a class of solutions of the Einstein field equations for such models. In [55], a system of second order differential equations is considered as an extension of geodesic equations and it is investigated by means of the KCC approach.
	
	The study of constant scalar curvature and constant flag curvature is continued in [51]; there, it is investigated the horizontal curvature of a pseudo-Finsler manifold. It is proven that, in the Berwald-Moor case, the horizontal and the flag curvature of the space vanish, while the induced curvatures on submanifolds are generally nontrivial.
	
	The investigation of rheonomic KCC models is continued in the work [55]; the autonomous case is extended to the rheonomic one by means of a geometrization of classical KCC theory on first order jet spaces. Here, the authors study the relations between spatial and time semisprays and define a nonlinear connection on the 1-jet space. They find the five invariants of the theory and point out the differences between the rheonomic case and the autonomous one, considering the geometric objects related to induced connections and KCC invariants.

\begin{small}

\end{small}
\end{document}